\documentclass[usenatbib]{mnras}

\usepackage{lscape}
\usepackage{graphicx}
\usepackage{amssymb}
\usepackage{amsmath}
\usepackage{url}
\usepackage{bm}
\usepackage{aas_macros}
\usepackage{rotating}

\usepackage{multirow}

\title[Intrinsic reddening of the Magellanic Clouds]{The intrinsic reddening of the Magellanic Clouds as traced by background galaxies -- II.
The Small Magellanic Cloud}

\author[C.~P.~M.~Bell et al.]{Cameron~P.~M.~Bell,${^1}$\thanks{E-mail:
  cbell@aip.de (CPMB)} Maria-Rosa L. Cioni,$^{1}$ A.~H.~Wright,$^{2}$ Stefano Rubele,$^{3,4}$
  \newauthor David~L.~Nidever,$^{5,6}$ Ben~L.~Tatton,$^{7}$ Jacco Th. van Loon,$^{7}$ Dennis Zaritsky,$^{8}$
  \newauthor Yumi Choi,$^{5,9}$ Samyaday Choudhury,$^{10,11}$ Gisella Clementini,$^{12}$
  \newauthor Richard de Grijs,$^{10,11,13}$ Valentin D. Ivanov,$^{14}$ Steven R. Majewski,$^{15}$
  \newauthor Marcella Marconi,$^{16}$ David Mart{\'i}nez-Delgado,$^{17}$ Pol Massana,$^{18}$ Ricardo R. Mu{\~n}oz,$^{19}$
  \newauthor Florian Niederhofer,$^{1}$ Noelia E. D. No{\"e}l,$^{18}$ Joana M. Oliveira,$^{7}$ Knut Olsen,$^{6}$
  \newauthor Clara M. Pennock,$^{7}$ V. Ripepi,$^{16}$ Smitha Subramanian,$^{20}$ A. Katherina Vivas$^{21}$\\
  $^{1}$Leibniz-Institut f{\"u}r Astrophysik Potsdam (AIP), An der Sternwarte 16, D-14482 Potsdam, Germany\\
  $^{2}$Astronomisches Institut, Ruhr-Universit{\"a}t Bochum, Universit{\"a}tsstr. 150, D-44801 Bochum, Germany\\
  $^{3}$Dipartimento di Fisica e Astronomia, Universit{\`a} di Padova, Vicolo dell'Osservatorio 2, I-35122 Padova, Italy\\
  $^{4}$INAF -- Osservatorio Astronomico di Padova, Vicolo dell'Osservatorio 5, I-35122 Padova, Italy\\
  $^{5}$Department of Physics, Montana State University, P.O. Box 173840, Bozeman, MT 59717, USA\\
  $^{6}$National Optical-Infrared Astronomy Research Laboratory (NOIRLab), 950 North Cherry Avenue, Tucson, AZ 85719, USA\\
  $^{7}$Lennard-Jones Laboratories, School of Chemical and Physical Sciences, Keele University, ST5 5BG, UK\\
  $^{8}$Steward Observatory, University of Arizona, 933 North Cherry Avenue, Tucson, AZ 85721, USA\\
  $^{9}$Space Telescope Science Institute, 3700 San Martin Drive, Baltimore, MD 21218, USA\\
  $^{10}$Department of Physics and Astronomy, Macquarie University, Balaclava Road, Sydney NSW 2109, Australia\\
  $^{11}$Research Centre for Astronomy, Astrophysics and Astrophotonics, Macquarie University, Balaclava Road, Sydney NSW 2109, Australia\\
  $^{12}$INAF -- Osservatorio di Astrofisica e Scienza dello Spazio di Bologna, Via Gobetti 93/3, I-40129 Bologna, Italy\\
  $^{13}$International Space Science Institute--Beijing, 1 Nanertiao, Zhongguancun, Hai Dian District, Beijing 100190, China\\
  $^{14}$European Southern Observatory, Karl-Schwarzschild-Str. 2, D-85748 Garching bei M{\"u}nchen, Germany\\
  $^{15}$Department of Astronomy, University of Virginia, Charlottesville, VA 22904, USA\\
  $^{16}$INAF -- Osservatorio Astronomico di Capodimonte, Via Moiariello 16, I-80131 Naples, Italy\\
  $^{17}$Instituto de Astrof{\'i}sica de Andaluc{\'i}a, CSIC, E-18080, Granada, Spain\\
  $^{18}$Department of Physics, University of Surrey, Guildford, GU2 7XH, UK\\
  $^{19}$Departamento de Astronom{\'i}a, Universidad de Chile, Camino del Observatorio 1515, Las Condes, Santiago, Chile\\
  $^{20}$Indian Institute of Astrophysics, Koramangala II Block, Bangalore, 560034, India\\
  $^{21}$Cerro Tololo Inter-American Observatory, National Optical Astronomy Observatory, Casilla 603, La Serena, Chile}
    
\begin{document}

\date{Accepted ?, Received ?; in original form ?}

\pagerange{\pageref{firstpage}--\pageref{lastpage}} \pubyear{2019}

\maketitle

\label{firstpage}

\begin{abstract}
We present a map of the total intrinsic reddening across $\simeq$\,34\,deg$^{2}$ of the Small Magellanic Cloud (SMC) derived using optical ($ugriz$) and
near-infrared (IR; $YJK_{\mathrm{s}}$) spectral energy distributions (SEDs) of background galaxies. The reddening map is created using a subsample of 29,274
galaxies with low levels of intrinsic reddening based on the \textsc{lephare} $\chi^{2}$ minimisation SED-fitting routine. We find statistically significant 
enhanced levels of reddening associated with the main body of the SMC compared with regions in the outskirts [$\Delta E(B-V)\simeq 0.3$\,mag]. A comparison
with literature reddening maps of the SMC shows that, after correcting for differences in the volume of the SMC sampled, there is good agreement between our results
and maps created using young stars. In contrast, we find significant discrepancies between our results and maps created using old stars or based on longer wavelength
far-IR dust emission that could stem from biased samples in the former and uncertainties in the far-IR emissivity and the optical properties of the dust grains in the latter.
This study represents one of the first large-scale categorisations of extragalactic sources behind the SMC and as such we provide the \textsc{lephare} outputs for our
full sample of $\sim$\,500,000 sources.
\end{abstract}

\begin{keywords}
Magellanic Clouds -- galaxies: photometry -- galaxies: ISM -- surveys -- dust, extinction
\end{keywords}

\section{Introduction}
\label{introduction}

The dust content of galaxies, and specifically the effects this has on photometric observations, is of great importance in the field of galaxy formation and evolution.
Within this framework, the Magellanic Clouds (MCs) are arguably two of the most studied galaxies, due primarily to their close proximity ($\sim 50-60$\,kpc) and wealth of different
stellar populations spanning all ages. The MCs are benchmark laboratories for studies of, amongst others, star formation at lower ($0.2-0.5$\,Z$_{\odot}$) metallicities
(e.g. \citealp*{Gouliermis14}; \citealp{Zivkov18}), constraining the extragalactic distance scale (and by extension the Hubble constant $H_{0}$, e.g.
\citealp*{deGrijs14}; \citealp{deGrijs15}; \citealp{Riess19,Freedman20}), as well as the metallicity dependence of the period--luminosity relation of variable
stars (see e.g. \citealp{Gieren18,Groenewegen18,Muraveva18,Ripepi19}). All such studies are dependent upon an understanding of the amount and spatial
distribution of dust across the MCs, however the issue is that the use of differently aged stellar populations to quantify the reddening in the MCs results in
statistically significant differences (this is further complicated by variations in metallicity, three-dimensional distributions of the stars and dust, etc.).
For example, the use of young stellar populations (e.g. young star clusters and Cepheids, \citealp{Nayak18,Joshi19}) results in reddening maps with enhanced levels of
reddening associated with regions that are actively forming stars and lower levels of reddening in more quiescent regions. In contrast, reddening maps based on older stars
[e.g. red clump (RC) and RR Lyrae] imply systematically lower levels and/or negligible homogeneously distributed levels of reddening (see e.g.
\citealp{Zaritsky02,Muraveva18,Gorski20}). Although these differences can be small, it is worth stressing that even a modest change in $\Delta E(B-V)$, such as the
difference between the RC reddening maps of \cite*{Haschke11} and \cite{Gorski20} of $\Delta E(B-V)=0.06$\,mag, can have significant implications. Even though both
maps use the same data from the third phase of the Optical Gravitational Lensing Experiment (OGLE-III; \citealp{Udalski03}) as an input, the difference between their
determined reddening values affects the inferred value for the Hubble constant at greater than the 3\,km\,s$^{-1}$\,Mpc$^{-1}$ level (see e.g. section~3.1 of
\citealp{Riess09}) and as well as the three-dimensional structure inferred from the use of resolved stars (e.g. \citealp{Ripepi17,Choi18}).

In \citet[hereafter Paper~I]{Bell19} we introduced a technique to map the intrinsic reddening of a foreground extinguishing medium using the spectral energy
distributions (SEDs) of background galaxies. This technique not only removes the aforementioned stellar population dependency on the derived reddening values,
but also probes the total reddening by sampling the full column depth of the extinguishing medium.
We piloted this method on two 1.77\,deg$^{2}$ regions in the Small Magellanic Cloud (SMC) -- the main body and the southern outskirts -- and demonstrated that when
using galaxies with low intrinsic reddening one can clearly see signs of higher intrinsic reddening associated with the main body compared with the more quiescent
outskirts. In this study, we extend this pilot study to map the total intrinsic reddening across the $\simeq$\,34\,deg$^{2}$ of the SMC covered by the combined
footprints of the Survey of the MAgellanic Stellar History (SMASH; \citealp{Nidever17}) and the Visible and Infrared Survey Telescope for Astronomy (VISTA) survey
of the Magellanic Clouds system (VMC; \citealp{Cioni11}).

The structure of this paper is as follows. In Section~\ref{creating_fitting_seds_of_galaxies} we describe the processes involved in creating and fitting the SEDs of
background galaxies as well as provide a comparison of the results of the SED fitting with other extragalactic studies. The details of how we create our intrinsic
reddening map of the SMC is presented in Section~\ref{determining_intrinsic_reddening_magellanic_clouds}. We discuss our reddening map and compare it with
literature reddening maps of the SMC based on different tracers in Section~\ref{discussion}, and summarise our findings in Section~\ref{summary}.

\section{Creating and fitting SEDs of galaxies}
\label{creating_fitting_seds_of_galaxies}

\subsection{Creating the SEDs}
\label{creating_the_seds}

Our data set consists of optical $ugriz$ and near-infrared (IR) $YJK_{\rm{s}}$ photometry taken as
part of the SMASH and VMC surveys,
respectively, that cover the wavelength range 0.3--2.5\,$\mu$m.
Paper~I describes in detail the processes involved in selecting likely background galaxies on which
to perform photometry. To briefly summarise, we use the VMC point-spread function (PSF)
photometric catalogues \citep{Rubele15,Rubele18} and retain those sources that satisfy the following colour--magnitude and
morphological selections: $J-K_{\rm{s}} > 1.0$\,mag, $K_{\rm{s}} > 15$\,mag, an associated stellar
probability of less than 0.33 and a $K_{\rm{s}}$-band sharpness index of greater than 0.5 (see Paper~I
for details regarding both the stellar probability and sharpness index).

Fluxes for each of our targets are measured using the Lambda Adaptive Multi-Band Deblending
Algorithm in R ({\sc lambdar}, v0.20.5; \citealt{Wright16}). The reader is directed to Paper~I for a
full discussion regarding the measurement of fluxes in addition to the calibration of both the optical
and near-IR data sets onto an AB magnitude system. We adopt the fluxes measured
using a default circular aperture of diameter 3\,arcsec.  Since Paper~I, we have optimised
the parameters used for per-exposure PSF estimation within \textsc{lambdar}\footnote{These updated PSF estimation
parameters are now used by default in v0.20.5.}, allowing for more accurate PSF estimation per observation.
This subsequently allows for more accurate (missed-flux) aperture corrections for our standard aperture photometry
(see \citealp{Wright16} for details) per observation, and more reliable photometry overall. 
The final flux measurement for a given source is
simply the weighted mean of all available measurements (including those in overlap regions) in that band and for which we set the weights
equal to the inverse square of the corresponding uncertainty on the flux.

\begin{figure}
\centering
\includegraphics[width=\columnwidth]{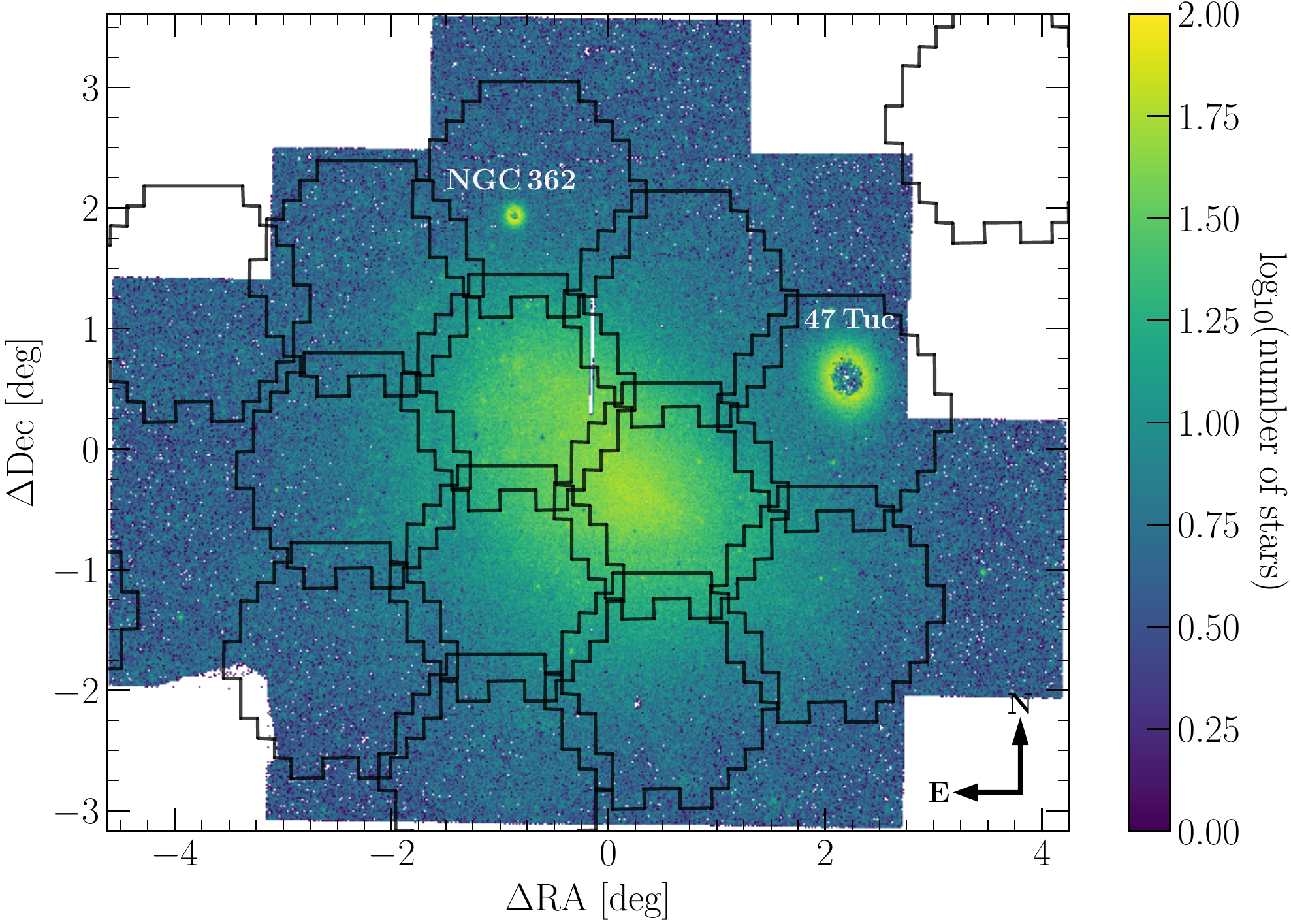}
\caption[]{Stellar density plot of the SMC as observed by the VMC survey. The origin of the plot corresponds to the
centre of the SMC as given by the HYPERLEDA catalogue of \protect\cite{Paturel03}. The black hexagons denote the positions of the SMASH fields.
The combined SMASH-VMC footprint is $\simeq$\,34\,deg$^{2}$.
The positions of the prominent foreground Milky Way globular clusters 47\,Tucanae and NGC\,362 are also shown. The vertical stripe at $\Delta$RA\,$\simeq-0.2$\,deg
is due to a narrow gap in the VMC observations that is currently being filled.}
\label{fig:smash_vmc_coverage}
\end{figure}

To ensure reliable fits to the
SEDs, we only retain sources for which we measure positive fluxes in at least four of the eight available
bands. Finally, we account for foreground Milky Way (MW) reddening by de-reddening the individual fluxes
by an amount equivalent to $E(B-V)=0.034$\,mag. This foreground subtraction is an average correction to the fluxes
as small-scale structures will cause variations across the survey area (see section~2.5 of Paper~I and \citealp{Bailey15} for details).
The resulting catalogue contains a total of 497,577 sources (hereafter referred to as the full SMC sample) distributed across $\simeq$\,34\,deg$^{2}$
(see Fig.~\ref{fig:smash_vmc_coverage}). Note that the
strange tile shape in the south-eastern VMC coverage is due to the presence of a Bridge tile that we do not include in this work as we will study the
reddening of the Magellanic Bridge in a future work.

\subsection{Fitting the SEDs}
\label{fitting_the_seds}

Paper~I provides a comprehensive discussion of the processes involved in fitting the SEDs of background galaxies,
including the choice of SED-fitting code, templates, redshift prior and attenuation curve, so we refer the reader to that description
for particulars.

Briefly, we adopt the \textsc{lephare} SED-fitting code\footnote{\url{http://cesam.lam.fr/lephare/lephare}} that allows one to fit several suites of templates
-- both theoretical and empirical -- including galaxies, quasi-stellar objects (QSOs) and stars to determine which
provides the best fit to the data based on a simple $\chi^{2}$-minimisation routine. In Paper~I we performed the SED
fits using both the theoretical and empirical galaxy templates to study the effects of template-dependency on the
resulting best-fitting reddening values. We found that the use of theoretical templates (in that case the so-called
COSMOS templates included as part of the \textsc{lephare} code) resulted in enhanced levels
of reddening at the $\Delta E(B-V)$\,$\simeq$\,0.2\,mag level compared to the reddening values inferred from the SED fits using
the empirical (AVEROI\_NEW) templates. After comparing the reddening maps inferred from background galaxies with
those available in the literature, based on various stellar components in the MCs, it was hard to reconcile
these enhanced levels of reddening with the difference in the depth of the SMC sampled by different tracers, thus
prompting us to prefer the reddening values based on the AVEROI\_NEW templates.

In this study, we only adopt
the empirical AVEROI\_NEW galaxy templates, in addition to the QSO and stellar templates discussed in Paper~I.
The redshift prior we adopt is included as part of \textsc{lephare} and has
been calibrated using the VIMOS VLT Deep Survey (VVDS; \citealp{LeFevre05}). Finally, we adopt the attenuation curve
resulting from the combined studies of \cite{Prevot84} and \cite{Bouchet85}\footnote{In Paper~I we only referenced \cite{Prevot84}, however the
attenuation curve provided in \textsc{lephare} covers the UV, optical and near-IR regimes and is in fact a combination of the UV attenuation
curve provided in \cite{Prevot84} and the optical/near-IR curve published in \cite{Bouchet85}. This is not explicitly clear from the \textsc{lephare} manual that
details the various attenuation curves provided as part of the code.} that is included as part of the \textsc{lephare} code.
As noted by \cite{Ilbert09}, there will inevitably be differences
in the attenuation curve from one galaxy to the
next, however the aforementioned curve is representative of the SMC attenuation curve and is also well suited
for the six galaxy types (E, Sbc, Scd, Irr and two starburst galaxies) from which the AVEROI\_NEW templates were created (see
Paper~I for details).

We run \textsc{lephare} on the standard aperture sample of galaxies allowing the redshift to vary from $z=0.0$ to 6.0 in steps
of $\Delta z=0.02$.  As discussed in Paper~I, the intervening dust in the SMC provides an additional component of the line-of-sight reddening. We
therefore allow additional reddening for all galaxy types, in contrast to the typical implementation for which additional
reddening is only permitted for galaxy templates of type Sc and bluer/later (see e.g. \citealp{Arnouts07}).
This allows us to use galaxies with spectral types of Sb and redder/earlier
as direct probes of the intrinsic reddening of the SMC. To limit degeneracies
in the best-fitting solutions, we allow the reddening to vary for the galaxy and QSO templates from $E(B-V)=0.0$ to 0.5\,mag
in steps of $\Delta E(B-V)=0.05$\,mag.
Table~\ref{tab:lephare_output} provides the \textsc{lephare} outputs, including photometric redshift, galaxy type,
best-fitting reddening values, etc. for the 497,577 sources in the full SMC sample and Table~\ref{tab:lephare_class} provides the numbers of sources
classified as galaxies, QSOs and stars based on the associated minimum $\chi^{2}$ value of the best-fitting template.

\begin{table*}
\caption[]{A sample of the \textsc{lephare} output for the 497,577 sources in the full SMC sample. We only show the ID, RA and Dec. (J2000.0), the best-fitting
photometric redshift with associated $1\sigma$ limits, the maximum likelihood photometric redshift with associated $1\sigma$ limits, the best-fitting galaxy/QSO template
and the associated $\chi^{2}$ value for the best-fitting galaxy/QSO template as a representation of its content. The full table (available as Supporting Information in the online
version of the paper) also includes the best-fitting $E(B-V)$ value, the best-fitting distance modulus, the number of photometric bands in the SED, the combination of
photometric bands in the SED (termed ``Context'' in \textsc{lephare}; see Notes below), the best-fitting stellar template and the associated
$\chi^{2}$ value for the best-fitting stellar template.}
\begin{tabular}{l c c c c c c c c c c}
\hline
ID   &   RA (J2000.0)   &   Dec. (J2000.0)   &   $z_{\rm{BEST}}$   &   $z_{\rm{BEST}}^{-1\sigma}$   &   $z_{\rm{BEST}}^{+1\sigma}$   &   $z_{\rm{ML}}$   &   $z_{\rm{ML}}^{-1\sigma}$   &   $z_{\rm{ML}}^{+1\sigma}$   &   Template$_{\rm{G/Q}}^{a}$   &   $\chi^{2}_{\rm{G/Q}}$\\
   &   deg   &   deg   &   &   &   &   &   &   &   &\\
\hline
1   &   8.22274   &   $-75.59664$   &   1.0984   &   1.0463   &   1.1526   &   0.6316   &   0.5288   &   1.1082   &   3   &   0.1229\\
2   &   8.23561   &   $-75.53493$   &   1.0176   &   0.1028   &   2.0265   &   0.6598   &   0.1635   &   1.1510   &   72   &   0.1256\\
3   &   8.24278   &   $-75.51370$   &   1.0178   &   0.0547   &   1.0429   &   0.0773   &   0.0309   &   0.1376   &   3   &   3.2894\\
4   &   8.17035   &   $-75.51329$   &   0.3901   &   0.3798   &   0.4001   &   --99.0   &   --99.0   &   --99.0   &   4   &   2064.5100\\
5   &   8.16404   &   $-75.51295$   &   1.0671   &   1.0416   &   1.1145   &   0.7988   &   0.3493   &   1.0796   &   6   &   0.7664\\
6   &   8.17099   &   $-75.51268$   &   1.0634   &   0.5162   &   1.0801   &   0.7641   &   0.5511   &   0.9438   &   71   &   0.4137\\
7   &   8.16723   &   $-75.51165$   &   1.0804   &   1.0607   &   1.1003   &   0.5580   &   0.5093   &   1.0875   &   4   &   14.5201\\
8   &   8.04942   &   $-75.50990$   &   1.0634   &   1.0513   &   1.0776   &   0.5704   &   0.4296   &   1.1121   &   4   &   5.2806\\
9   &   8.20849   &   $-75.50972$   &   0.5221   &   0.5114   &   0.5331   &   0.5234   &   0.5036   &   0.5448   &   2   &   12.5986\\
10   &   7.88062   &   $-75.50932$   &   0.1384   &   0.1189   &   0.1593   &   0.1418   &   0.1215   &   0.1618   &   48   &   19.8340\\
\hline
\end{tabular}
\vspace{1pt}
\begin{flushleft}
Notes: The reader is referred to the \textsc{lephare} documentation for more details regarding the outputs of the code as well as the various template libraries contained within.\\
$^{a}$ Best-fitting galaxy/QSO templates are as follows: (1) Seyfert 1.8, (2) Seyfert 2, (3) QSO1, (4) BQSO1, (5) TQSO1, (6) QSO2, (7) Torus (QSO2), (8) Mrk 231,
(9) IRAS 19254--7245, (10) NGC 6240, (11--31) E, (32--47) Sbc, (48--58) Scd, (59--68) Irr, (69--72) Starburst.\\
$^{b}$ Context is a numerical representation in \textsc{lephare} specifying the combination of bands present in the input catalogue and is defined as
$\sum_{i=1}^{i=N} 2^{i-1}$, where $i$ is the band number (in our case $u=1$, $g=2$, ..., $J=7$, and $K_{\rm{s}}=8$), and $N$ is the total number of bands.
\end{flushleft}
\label{tab:lephare_output}
\end{table*}

\begin{table}
\caption[]{Number of sources classified as a galaxy, QSO or star by \textsc{lephare}.
The values in parentheses denote the corresponding percentages of the total
number of sources in the full SMC sample (see Section~\ref{fitting_the_seds}).}
\begin{tabular}{c c c}
\hline
Galaxies   &   QSOs   &   Stars\\
\hline
190,310 (38.2)   &   271,556 (54.6)   &   35,711 (7.2)\\
\hline
\end{tabular}
\label{tab:lephare_class}
\end{table}

\subsection{Galaxy photometric redshift and spectral type distributions}
\label{galaxy_photo_z_spectral_type_distribution}

\begin{figure}
\centering
\includegraphics[width=\columnwidth]{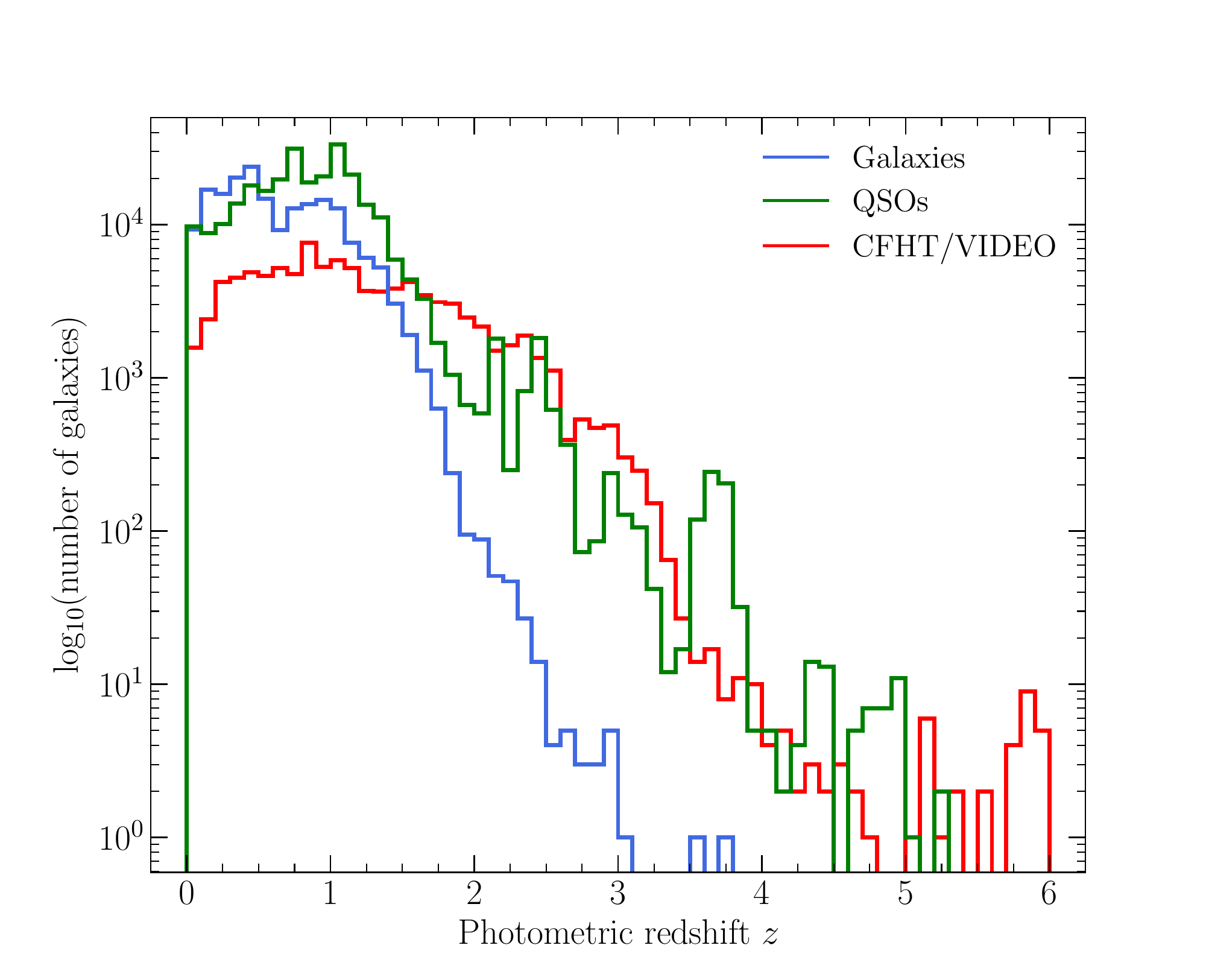}
\caption[]{Photometric redshift distributions of sources classified by \textsc{lephare} as galaxies (blue) and QSOs (green) in the full SMC sample. For
comparison, we also show the redshift distribution of galaxies in the CFHT/VIDEO (red) sample (see text for details).}
\label{fig:redshift_dist_full_SMC}
\end{figure}

\begin{figure}
\centering
\includegraphics[width=\columnwidth]{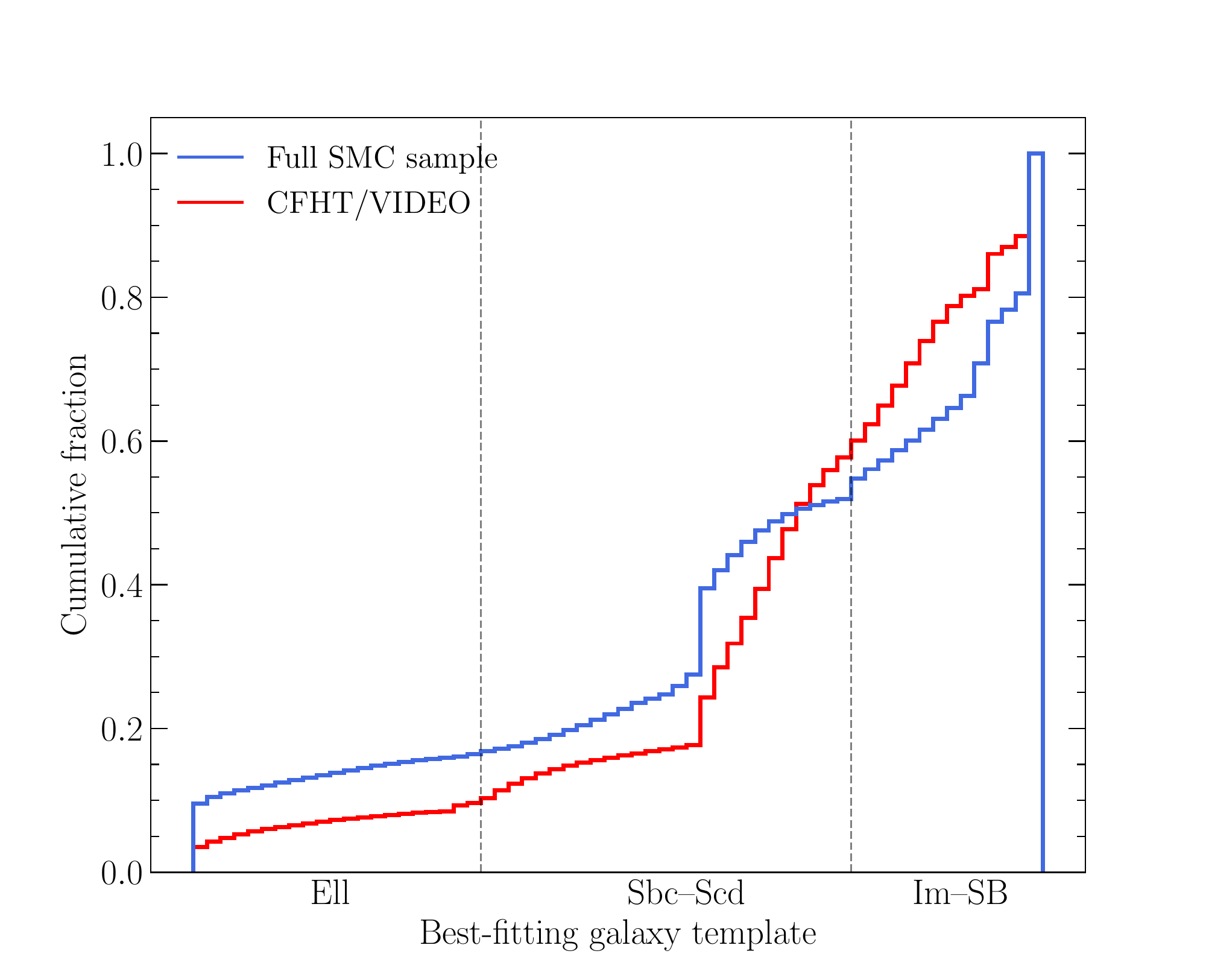}
\caption[]{Normalized cumulative distributions showing the distribution of best-fitting galaxy templates (elliptical, spiral and irregular/starburst) in the full
SMC (blue) and CFHT/VIDEO (red) samples.}
\label{fig:best_model_dist_full_SMC}
\end{figure}

Given that this study represents one of the first large-scale homogeneous cataloguing and categorisations of background galaxies behind the SMC, it is
relevant to place these results in context by discussing some galaxy statistics. Figs.~\ref{fig:redshift_dist_full_SMC} and \ref{fig:best_model_dist_full_SMC}
show the photometric redshift and best-fitting galaxy template distributions, respectively, for the full SMC sample compared with those resulting from the analysis
of the combined CFHTLS-DF1 $ugriz$ and VISTA Deep Extragalactic Observations (VIDEO) $ZYJHK_{\mathrm{s}}$ data performed by \cite{Jarvis13}.

We preferentially adopt the photometric redshift from the median of the maximum likelihood distribution (termed $z_{\mathrm{ML}}$ in \textsc{lephare}).
When this is not available, however, we simply adopt the photometric redshift corresponding to the redshift with the minimum $\chi^{2}$ value (termed $z_{\mathrm{BEST}}$;
see Section 3.5 of \citealp{Ilbert09} for details). Our reasoning for preferentially adopting $z_{\mathrm{ML}}$ is that, as discussed in \cite{Ilbert13},
whilst $z_{\mathrm{ML}}$ and $z_{\mathrm{BEST}}$ are broadly similar, the use of the median of the likelihood distribution results in more reliable uncertainties
on the photometric redshift and also reduces the effect of aliasing in photometric redshift space\footnote{Note that despite having increased the uncertainties on the
individual \textsc{lambdar} fluxes by a factor of 1.5 as well as including an additional systematic uncertainty of 0.1\,mag in each of our eight bands, we find that for
$\simeq$\,16 per cent of the sources classified as galaxies/QSOs in the full SMC sample the associated 1$\sigma$ limits on $z_{\mathrm{BEST}}$ are unphysical i.e.
$z_{\mathrm{BEST}}^{-1\sigma} > z_{\mathrm{BEST}}$ or $z_{\mathrm{BEST}}^{+1\sigma} < z_{\mathrm{BEST}}$. As stated in \cite{Ilbert09}, inflating the uncertainties
on the individual fluxes does not have an impact on the derived value for $z_{\mathrm{BEST}}$, but only acts to broaden the $\chi^{2}$ distribution and hence redshift
uncertainty. Thus, even if we were to further inflate the flux uncertainties the photometric redshift distribution shown in Fig.~\ref{fig:redshift_dist_full_SMC} would not
change.}. In Paper~I
we found clear evidence (see their figs.~10 and 11) of differences in the peak of the galaxy redshift distribution between the two regions studied (the outskirts and the
south-western main body), that we attributed to the combined effects of crowding and incompleteness in the latter. For the full SMC sample, we calculate a median redshift
of $z_{\mathrm{med}}=0.55$ and a median absolute deviation of $z_{\mathrm{MAD}}=0.44$. At higher redshifts, we see a steep decline with only
three galaxies at redshifts greater than 3. In Fig.~\ref{fig:redshift_dist_full_SMC} we also show the QSO redshift distribution of the
full SMC sample, which are those sources with a lower $\chi^{2}$ value for a QSO template than for a galaxy template. Given that, on average, we expect QSOs to be
brighter than the quiescent background galaxy population, it is no surprise to see the QSO redshift distribution peak and extend to higher redshifts ($z_{\mathrm{med}}=0.84$;
$\simeq$\,1000 sources with $z > 3$)\footnote{As noted in Paper~I, it can be difficult to reliable differentiate between galaxies and QSOs based only on the $\chi^{2}$
values returned by \textsc{lephare} as this can result in a significant number of normal galaxies being classified as QSOs. Thus, the median and spread of the galaxy
and QSOs redshift distributions, as well as the number of classified galaxies and QSOs listed in Table~\ref{tab:lephare_class}, should be treated
with care.}. Comparing the galaxy redshift distribution of the full SMC sample with that of the CFHT/VIDEO sample, we see that the latter not
only has a lower peak, but a broader distribution than that of the full SMC sample ($z_{\mathrm{med}}=1.06$; $z_{\mathrm{MAD}}=0.72$).
This discrepancy can be attributed to the combination of differences in both the areal coverage
and photometric sensitivity of the two samples. Whilst the CFHT/VIDEO photometry is much deeper than our combined SMASH-VMC photometry [with 5$\sigma$
magnitude limits $\gtrsim$\,1.5\,mag (AB) across all eight bands], it only covers an area of 1\,deg$^{2}$ on the sky compared to the $\simeq$\,34\,deg$^{2}$ covered by
our sample. As regards the distribution of best-fitting galaxy templates, Fig.~\ref{fig:best_model_dist_full_SMC} shows that the distributions of the full SMC
and CFHT/VIDEO samples are consistent at the 10--15 per cent level across all spectral types.

\subsection{Comparison of photometric to spectroscopic redshifts}
\label{comparison_photo_spectro_z}

\begin{figure}
\centering
\includegraphics[width=\columnwidth]{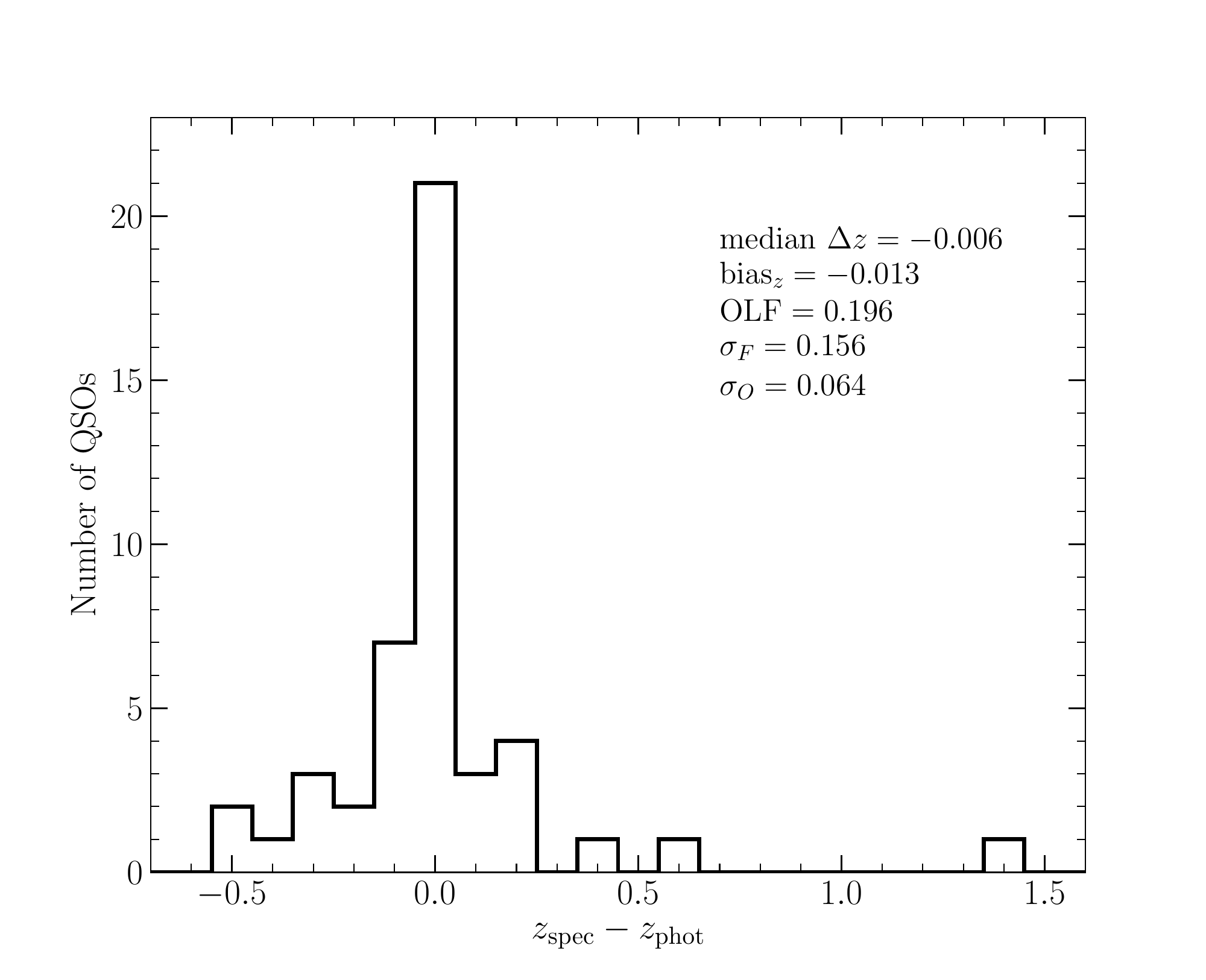}
\caption[]{Histogram showing the differences between the spectroscopically determined redshifts of QSOs behind the SMC and the corresponding photometric
redshifts calculated by \textsc{lephare}. The various statistics regarding the comparison of the two redshift determinations are discussed in the text.}
\label{fig:redshift_comp}
\end{figure}

\begin{table}
\caption[]{Comparison between the spectroscopically determined and calculated photometric redshifts for QSOs behind the SMC. The full table is available as
Supporting Information in the online version of the paper.}
\begin{tabular}{l c c c}
\hline
Name   &   $z_{\mathrm{spec}}$   &   $z_{\mathrm{phot}}$   &   Ref.\\
\hline
XMMU J003917.7--730330   &   0.345   &   $0.400\pm0.014$   &   1\\
MQS J004143.75--731017.1   &   0.217   &   $0.259^{+0.026}_{-0.029}$   &   2\\
2MASS J00414502--7254356   &   0.267   &   $0.518^{+0.015}_{-0.014}$   &   3\\
MQS J004152.35--735626.8   &   0.422   &   $0.289^{+0.065}_{-0.019}$   &   2\\
MQS J004336.54--735615.1   &   0.661   &   $0.279\pm0.014$   &   2\\
MQS J004736.12-724538.2   &   0.572   &   $0.614^{+0.017}_{-0.014}$   &   3\\
OGLE SMC100.4 26477   &   0.288   &   $0.264^{+0.024}_{-0.022}$   &   3\\
MQS J005311.80-740852.6   &   0.293   &   $0.258^{+0.076}_{-0.089}$   &   2\\
XMMU J005325.9-714821   &    0.982   &   $0.881^{+0.014}_{-0.015}$   &   1\\
OGLE SMC105.7 34076   &   0.505   &   $0.500\pm0.014$   &   3\\
\hline
\end{tabular}
\vspace{1pt}
\begin{flushleft}
Notes: $^{a}$ Associated 1$\sigma$ limits on $z_{\mathrm{BEST}}$ are unphysical (see footnote 4 in text).\\
References: (1) \protect\cite{Maitra19}, (2) \protect\cite{Kozlowski13}, (3) \protect\cite{Kozlowski11},
(4) Public ESO Spectroscopic Survey of Transient Objects (PESSTO; \protect\citealp{Smartt15}).
\end{flushleft}
\label{tab:spec_phot_z}
\end{table}

To further test the validity of the \textsc{lephare} outputs we compare the calculated photometric redshifts to spectroscopically determined redshifts of QSOs behind the SMC.
Our sample comprises 232 spectroscopically confirmed QSOs from \cite{Dobrzycki03a,Dobrzycki03b}, \cite{Geha03}, \cite{Veron-Cetty10},
\cite*{Kozlowski11}, \cite{Kozlowski13}, \cite{Ivanov16}, and \cite{Maitra19}, for which we find 229 unique matches in the VMC PSF photometric catalogues
covering the SMC within a 1\,arcsec matching radius. Of the 229 QSOs, only 46 satisfy the colour--magnitude and morphological selections
described in Section~\ref{creating_the_seds}. Table~\ref{tab:spec_phot_z} lists the comparison between the photometric redshifts calculated by \textsc{lephare}
and the corresponding spectroscopically determined redshifts. Fig.~\ref{fig:redshift_comp} shows that the two redshift determinations agree, in general, very
well with a median difference of $\Delta z=-0.006$ and median absolute deviation of $\Delta z=0.101$.
Following \cite{Dahlen13}, we also report additional statistics including the bias, outlier fraction (OLF), full
scatter ($\sigma_{F}$) and scatter having removed outliers ($\sigma_{O}$)\footnote{bias$_{z} = \mathrm{mean}[\Delta z/(1+z_\mathrm{spec})]$; OLF is
the fraction of sources that are outliers, defined as $|\Delta z|/(1+z_\mathrm{spec}) > 0.15$; $\sigma_{F} = \mathrm{rms}[\Delta z/(1+z_\mathrm{spec})]$;
$\sigma_{O}$ is the same as $\sigma_{F}$, only calculated having removed sources classified as outliers.}. Even though the sample from which these
statistics are drawn is small, it is worth noting that they are all consistent with the statistics calculated by \cite{Dahlen13}
who tested several photometric redshift codes on a sample of $\sim$\,600 galaxies from the Cosmic Assembly Near-infrared Deep Extragalactic
Legacy Survey (CANDELS) GOODS-S data (see their tables~2 and 3).

\section{Determining the intrinsic reddening}
\label{determining_intrinsic_reddening_magellanic_clouds}

In Paper~I we introduced several effects that could potentially affect the inferred reddening values, including the number of
bands used in the SED fit, blending and incompleteness at fainter magnitudes and redshift probability distributions (PDF$z$)
with multiple peaks (see appendix~A of Paper~I for details). To briefly summarise, we retained only those sources for which
fluxes in all eight bands are available, those sources that are brighter than the magnitude at which incompleteness starts to
affect the sample and those sources that only have a single peak in the PDF$z$. We found that by imposing these constraints
on the sample, the inferred reddening maps were not significantly affected. Although there were minor morphological differences,
the median reddening across a tile was only affected at the $\Delta E(B-V)$\,$\simeq$\,0.03\,mag level. Note that the magnitude cut imposed
as a result of decreasing completeness is performed on a VMC tile-by-tile basis. Imposing these constraints reduces our sample size from
the 497,577 sources in the full SMC sample to 197,821 galaxies/QSOs in the cleaned SMC sample.

\begin{figure}
\centering
\includegraphics[width=\columnwidth]{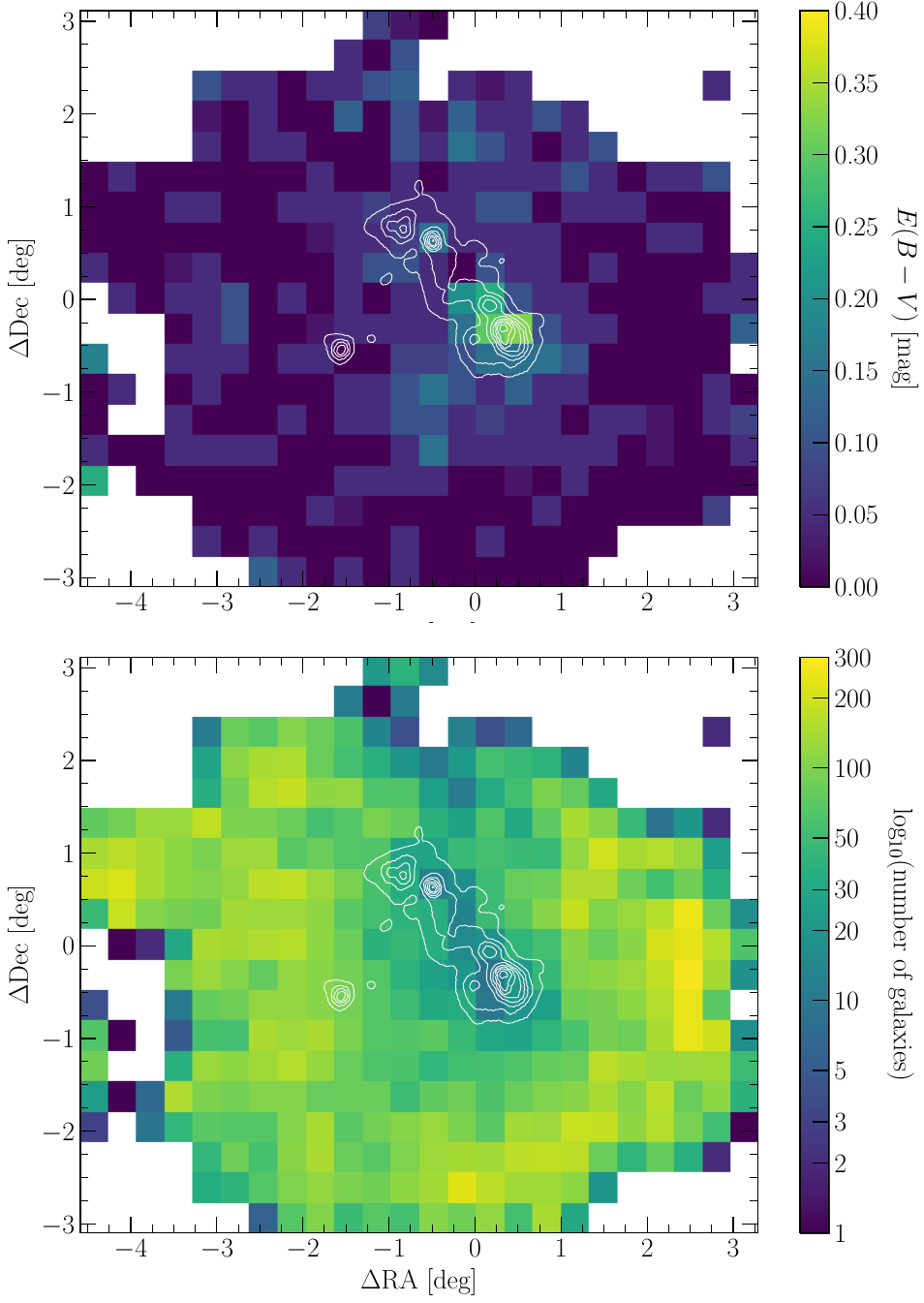}
\caption[]{\emph{Top panel:} $20\times20$\,arcmin$^{2}$ resolution reddening map of the combined SMASH-VMC footprint covering $\simeq$\,34\,deg$^{2}$ of the SMC.
This map has been created using only galaxies with spectral types of Sbc and redder/earlier according to the AVEROI\_NEW templates and that satisfy the three
criteria designed to refine the sample (see Section~\ref{fitting_the_seds}). \emph{Bottom panel:} As the top panel, but showing the number density of galaxies used
to create the reddening map. The contours in both panels represent the IRAS 100\,$\mu$m dust emission.}
\label{fig:red_map_num_gals_full_SMC}
\end{figure}

Fig.~\ref{fig:red_map_num_gals_full_SMC} shows a $20\times20$\,arcmin$^{2}$ resolution reddening map covering $\simeq$\,34\,deg$^{2}$ of the SMC.
The $E(B-V)$ value associated with a given bin is simply the median value of all best-fitting $E(B-V)$ values. The total number
of galaxies used to create the reddening map (with spectral types of Sb and redder/earlier from the cleaned SMC sample) is 29,274 distributed across 343 bins.
Note that as demonstrated in Paper~I (see also Section~\ref{fitting_the_seds}), reddening maps based on QSOs and/or all galaxies are inconsistent with the
morphology of dust within the SMC, and thus we use only those galaxies with low levels of intrinsic reddening. The number of galaxies in a given bin ranges from 1 to 276, with a
median of 82. There are 5 bins for which the associated $E(B-V)$ value is based on only a single galaxy. Several bins in the outskirts of the combined SMASH-VMC
footprint contain only a few galaxies and these likely explain the enhanced levels of reddening associated with some outskirt bins. In the central regions of the SMC,
that are affected by incompleteness due to crowding, we note that the number of galaxies per bin is significantly lower
than the median value of 82 (see bottom panel of Fig.~\ref{fig:red_map_num_gals_full_SMC}). However, in the regions of Fig.~\ref{fig:red_map_num_gals_full_SMC}
traced by the IRAS 100\,$\mu$m dust emission contours, we have at least 9 galaxies in all bins.

From Fig.~\ref{fig:red_map_num_gals_full_SMC} we can clearly see the enhanced levels of intrinsic reddening associated with the main body, and in particular the
south-western regions. Whilst we detect increased levels of intrinsic reddening in some north-eastern regions of the main body, there is no clear continuous
region of enhanced reddening along the length of the main body (as traced by the 100\,$\mu$m emission). This pattern is remarkably similar to the 5780 and 5797\,$\rm{\AA}$
diffuse interstellar band (DIB) maps produced by \citet[see their fig. 7]{Bailey15}. Furthermore, we detect no clear signature of enhanced levels
of intrinsic reddening, with respect to the low-level background, in the region associated with the Wing ($\Delta$RA$\,\simeq$\,--1.5, $\Delta$Dec$\,\simeq$\,--0.5\,deg).
In contrast, Fig.~\ref{fig:red_map_num_gals_full_SMC} shows small regions of enhanced levels of intrinsic reddening to both the north and south of the main body. These regions do
not coincide with any known morphological features (see e.g. \citealp{ElYoussoufi19}), however the regions to the south do coincide with overabundances of
Ca\,{\footnotesize{II}} K and Na\,{\footnotesize{I}} D$_{2}$ atomic absorption as measured by \cite{Bailey15}.
The median intrinsic reddening across the combined SMASH-VMC footprint is $E(B-V)_{\mathrm{med}}=0.05$\,mag, with a maximum value of $E(B-V)=0.33$\,mag
in the south-western region of the main body. The median uncertainty on the reddening within a given bin (calculated as the standard deviation of the best-fitting reddening
values within a given bin) is $\sigma_{E(B-V)_{\mathrm{med}}}=0.10$\,mag. For bins with only a single galaxy we assume an uncertainty of 0.05\,mag based on the adopted
spacing of $E(B-V)=0.05$\,mag in the \textsc{lephare} fitting (see Section~\ref{determining_intrinsic_reddening_magellanic_clouds}). Table~\ref{tab:smc_red_bins}
provides the median reddening values, standard deviations and the number of galaxies in each of the 343 bins covering the combined SMASH-VMC footprint
of the SMC.

\begin{table*}
\caption[]{Median reddening values, standard deviations and the number of galaxies in each of the 343 bins covering the combined SMASH-VMC footprint
of the SMC. The full table is available as Supporting Information in the online version of the paper.}
\begin{tabular}{c c c c c}
\hline
RA (J2000.0)   &   Dec. (J2000.0)   &   $E(B-V)_{\mathrm{med}}$   &   $\sigma_{E(B-V)}$   &   No. galaxies\\
deg   &   deg   &   mag   &   mag   &\\
\hline
$-2.45490$   &   $-2.93296$   &   0.125   &   0.144   &   6\\
$-2.12704$   &   $-2.93296$   &   0.050   &   0.096   &   66\\
$-1.79918$   &   $-2.93296$   &   0.000   &   0.085   &   101\\
$-1.47132$   &   $-2.93296$   &   0.050   &   0.087   &   49\\
$-1.14346$   &   $-2.93296$   &   0.000   &   0.119   &   68\\
$-0.81560$   &   $-2.93296$   &   0.050   &   0.085   &   56\\
$-0.48774$   &   $-2.93296$   &   0.000   &   0.105   &   107\\
$-0.15988$   &   $-2.93296$   &   0.000   &   0.080   &   88\\
0.16798   &   $-2.93296$   &   0.000   &  0.099   &   119\\
0.49584   &   $-2.93296$   &   0.000   &   0.101   &   55\\
\hline
\end{tabular}
\label{tab:smc_red_bins}
\end{table*}

\section{Discussion and comparisons}
\label{discussion}

In this study, we have presented a map of the total intrinsic reddening covering $\simeq$\,34\,deg$^{2}$ of the SMC as defined by the combined SMASH-VMC footprint.
Following Paper~I, the use of background galaxies with low levels of intrinsic reddening allows us to observe clear signatures of enhanced levels of reddening
associated with the main body of the SMC. However, we do not see such enhancements in either the northern region of the main body or the SMC Wing. In this
section we will compare our reddening map with large-scale reddening maps of the SMC based on different tracers to highlight potential discrepancies
with respect to both the line-of-sight reddening values and the distribution of dust across the galaxy. Note that in the figures presented below
we do not apply any correction for the difference in the volume sampled between the reddening map based on background galaxies
and those based on the stellar components of the SMC. Furthermore, if the literature maps include the
contribution from the foreground MW, then we remove this component by simply subtracting $E(B-V)=0.034$\,mag (see Section~\ref{creating_the_seds})
from each reddening measurement. As a result of removing the MW component of the reddening, a small fraction of measurements in the literature maps become
negative. We therefore exclude such measurements from our analysis, but note that their removal does not affect the conclusions of the comparisons below.

\subsection{Literature SMC reddening maps}
\label{smc_reddening_maps_literature}

In Paper~I we introduced several literature reddening maps based on various stellar components of the SMC. The discussion, however, was limited to
only the two tiles investigated as part of that pilot study. In this study, we retain the reddening maps of \citet[specifically that based on the cool stars]{Zaritsky02},
\cite{Muraveva18}, \cite{Rubele18} and Tatton et al. (submitted), and refer the reader to Paper~I for details concerning each of these. Additionally, we include as
part of our comparison recent reddening maps produced by \cite{Joshi19} and \cite{Gorski20} as well as the dust emission map based on far-IR and submillimetre observations
taken as part of the \emph{HERschel} Inventory of The Agents of Galaxy Evolution (HERITAGE) project \citep{Meixner13}.

\subsubsection{\protect\cite{Joshi19}}
\label{joshi19}

\cite{Joshi19} used $V$- and $I$-band data from OGLE-IV \citep*{Udalski15} to
create a $\simeq$\,5\,deg$^{2}$ reddening map of the SMC using a combination of 4546 fundamental mode and first overtone mode Cepheids. The authors
do not derive reddening values for individual Cepheids, but instead use period--luminosity relations to determine the reddening towards a given region
containing a minimum of 10 Cepheids. In total, there are 136 regions that can be used to map the reddening across the SMC. To transform the derived
$E(V-I)$ values into $E(B-V)$ we adopt the prescription of \citet*{Cardelli89} [$E(V-I) = 1.32\times E(B-V)$].

\subsubsection{\protect\cite{Gorski20}}
\label{gorski20}

\cite{Gorski20} recently re-determined the reddening in the SMC using $V$- and $I$-band data from OGLE-III based on the colour of the RC. The data
used are identical to those in \cite{Haschke11}, however \cite{Gorski20} calculate the intrinsic colour for unreddened RC stars observationally using a number of tracers
including late-type eclipsing binaries, measurements of blue supergiants and B-type stars, determining a bluer intrinsic colour of $V-I=0.814$\,mag
(cf. $V-I=0.89$\,mag by \citealp{Haschke11}).
The reddening is then measured by calculating the difference between the observed and intrinsic colour of the RC in $3 \times 3$\,arcmin$^{2}$
bins across the $\simeq$\,14\,deg$^{2}$ OGLE-III footprint of the SMC. The publicly available reddening
maps\footnote{\url{https://araucaria.camk.edu.pl/index.php/magellanic-clouds-extinction-maps/}} list $E(B-V)$ instead of $E(V-I)$ and were converted by \cite{Gorski20}
using the same conversion as that noted in Section~\ref{joshi19}.

\subsubsection{\protect\cite{Gordon14}}
\label{gordon14}

As part of the HERITAGE project, \cite{Gordon14} used photometric data in five bands from 100 to 500\,$\mu$m to study the submillimetre excess in the
SMC across an area of $\simeq$\,37\,deg$^{2}$. \cite{Gordon14} employ simple dust emission models based on one or two modified blackbodies, and of the
three different models tested, we adopt the results from the use of the so-called BEMBB model (a single temperature blackbody modified
by a broken power-law emissivity) as this model provides the lowest residuals (see e.g. \citealp{Gordon14,Roman-Duval14})\footnote{The BEMBB model
is clearly stated as the recommended model on K. Gordon's website (\url{https://karllark.github.io/data_magclouds_dustmaps.html}) which provides the
results of all three models tested.}. For each of the tested models, \cite{Gordon14} provide three different realisations of the
best-fitting parameters. Of these, we adopt the best-fitting parameters corresponding to the expectation value of the full likelihood function for each parameter
(see section 4.3 of \citealp{Gordon14} for details).

The most important parameter of the BEMBB model, in terms of investigating the dust content across the SMC, is the dust surface density $\Sigma_{\mathrm{d}}$.
To transform the dust surface densities to $E(B-V)$ values for a direct comparison with our reddening map based on the background
galaxies, we follow the formalism described by \cite{Whittet03}. The relation between the total extinction at a given wavelength, $\lambda$, and the optical
depth of extinction caused by the dust, $\tau_{\mathrm{d}}$, is given by:

\begin{equation}
A_{\lambda} = 1.086\tau_{\mathrm{d}}.
\end{equation}

\noindent We can express this total extinction in terms of the extinction efficiency factor, $Q_{\mathrm{ext}}$:

\begin{equation}
A_{\lambda} = 1.086 N_{\mathrm{d}} \pi a^{2} Q_{\mathrm{ext}},
\end{equation}

\noindent where $N_{\mathrm{d}}$ is the column density of the dust and $a$ is the radius of the dust grains (see section 3.1.1 of \citealp{Whittet03} for a full
derivation). The dust surface density $\Sigma_{\mathrm{d}}$ is related to $N_{\mathrm{d}}$ via:

\begin{equation}
\Sigma_{\mathrm{d}} = \frac{4}{3} \pi a^{3} N_{\mathrm{d}} \rho_{\mathrm{d}},
\end{equation}

\noindent where $\rho_{\mathrm{d}}$ is the density of the dust grains. We can calculate the $V$-band extinction using the $V$-band extinction
efficiency factor $Q_{V}$, such that:

\begin{equation}
A_{V} = 0.8145 \frac{\Sigma_{\mathrm{d}} Q_{V}}{a \rho_{\mathrm{d}}}.
\label{eqn:av_sigma_dust}
\end{equation}

\noindent We adopt the following values for the remainder of the variables in Eqn.~(\ref{eqn:av_sigma_dust}): $Q_{V}=1.4$ \citep{Chlewicki85}
and, assuming ``classical'' dust grains, $a=0.1$\,$\mu$m and $\rho_{\mathrm{d}}=3$\,g\,cm$^{-3}$ \citep{Mann00,Whittet03,Kohout14}.
Finally, we convert $A_V$ into $E(B-V)$ following the standard $A_V = 3.1 \times E(B-V)$ \citep{Cardelli89}. Note that the values of $\Sigma_{\mathrm{d}}$
represent only the internal dust content of the SMC. Prior to fitting the HERITAGE data, \cite{Gordon14} use the integrated MW H\,{\footnotesize{I}} velocity
maps in the directions of the SMC \citep{Stanimirovic99,Muller03} to subtract the structured emission due to the MW.

\subsection{Quantitative comparisons}
\label{quantitative_comparisons}

\begin{figure*}
\centering
\includegraphics[width=\textwidth]{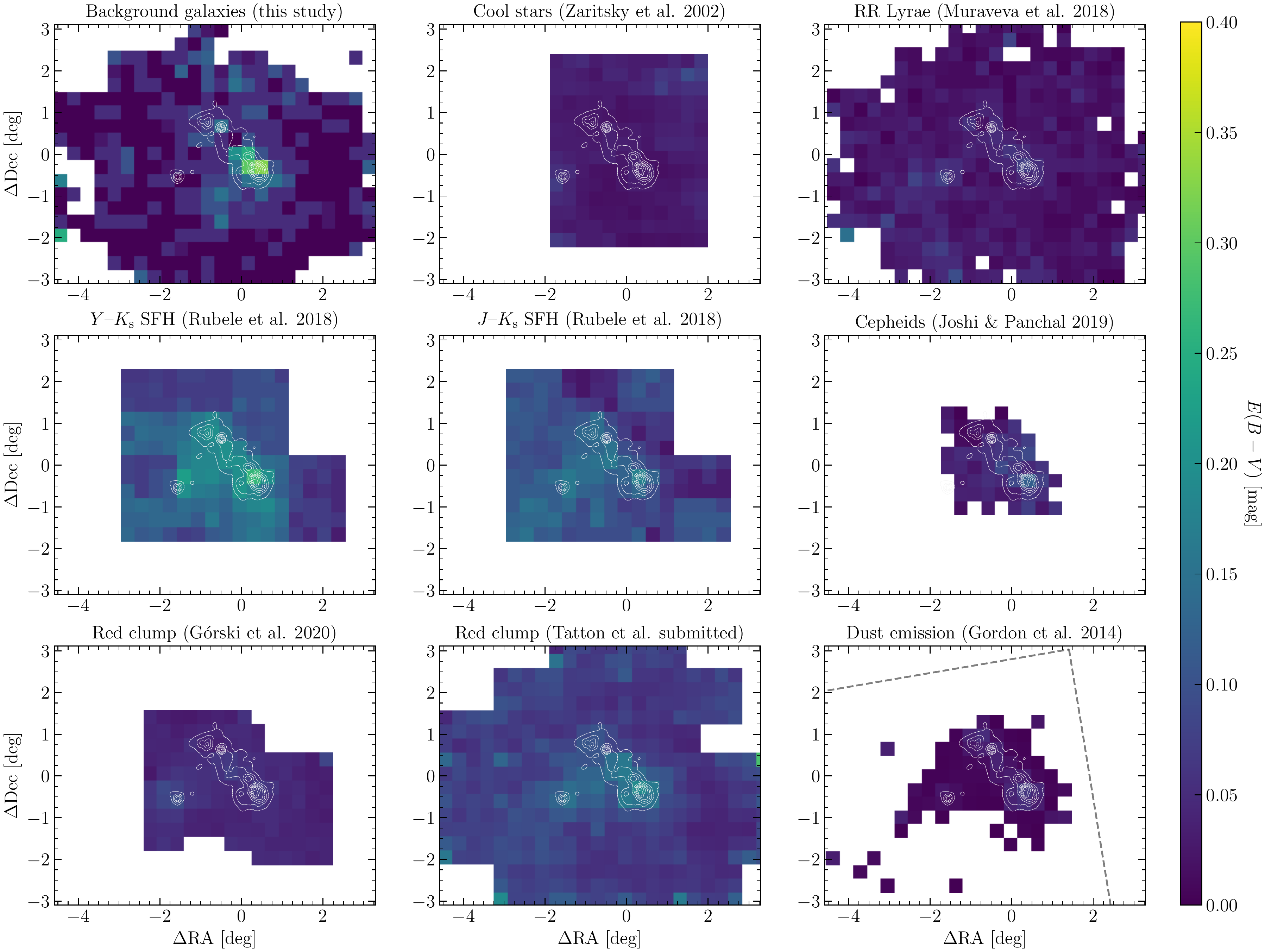}
\caption[]{$20 \times 20$\,arcmin$^{2}$ resolution reddening maps covering the SMC. Each panel refers to a different literature
source. The contours in all panels represent the IRAS 100\,$\mu$m dust emission. The dashed line in the lower right panel represents the footprint
of the HERITAGE observations.}
\label{fig:red_comp_single}
\end{figure*}

Fig.~\ref{fig:red_comp_single} presents the comparison between our reddening map based on galaxies with low levels of intrinsic reddening and the various reddening
maps of the SMC discussed in Section~\ref{smc_reddening_maps_literature}. To facilitate the comparison, we have resampled the literature maps to a
$20 \times 20$\,arcmin$^{2}$ resolution by simply taking the median reddening value of all stars/cells that fall within a given bin. This resampling of the literature maps
will undoubtedly result in the loss of small-scale (less than 20\,arcmin in size) features in these maps. Any large-scale morphological features, however, such as pervasive
enhancements of dust along the main body of the SMC, should be preserved and visible in the maps presented in Fig.~\ref{fig:red_comp_single}.

Of the eight literature maps shown, three (\citealp{Zaritsky02}, \citealp{Muraveva18} and \citealp{Gorski20}) show no obvious signs of enhanced levels of
reddening associated with the main body of the SMC, when compared with the levels of reddening in regions away from the main body.
In contrast, the remaining five maps (\citealp{Gordon14}, the two realisations of
\citealp{Rubele18}, \citealp{Joshi19} and Tatton et al. submitted) all exhibit clear signs of enhanced reddening in the south-western region of the main body
($\Delta$RA$\,\simeq$\,0.35, $\Delta$Dec$\,\simeq$\,--0.35\,deg) as well as other regions in some cases.

The presence of enhanced levels of dust appears
to be tied to the type of star one uses to measure the reddening. The maps that show no significant levels of reddening across the SMC are all based on
intermediate-age/old stars, whereas those that do show higher levels of intrinsic reddening (barring the \citealp{Gordon14} map)
are either based on or include young stars. This difference in the reddening values based on old/young stellar populations can be ascribed to the spatial distribution of
each population. The older stars are more uniformly distributed within the SMC as opposed to the young stars which tend to be concentrated in dense regions associated
with recent star formation, i.e., their birth sites. Thus, in contrast to these young stars, the reddening values based on old stars will be more reflective of the diffuse
interstellar medium (ISM) which,
coupled with the averaging over spatial regions, means that dense regions with small filling factors will effectively be missed by the spatially averaged measurements
from the older stars. Note that this effect is also observed in the two reddening maps presented by \cite{Zaritsky02} based on the young, densely concentrated hot stars and the old,
uniformly distributed cool stars.

Although the reddening maps based on older stellar populations tend to show very low levels of reddening across the SMC, the near-IR RC map of Tatton et al.
(submitted) is a notable exception, despite the similar technique adopted to compute all RC reddening maps (see also \citealp{Haschke11,Gorski20}).
In Paper~I (see section~5.2.2) we discussed some of the potential differences between the use of RC stars in the optical and near-IR regimes. The main difference
between the maps shown in Fig.~\ref{fig:red_comp_single}, however, is that Tatton et al. (submitted) determine reddening values for individual stars as opposed
to for a discrete cell containing numerous stars. The use of cells as opposed to individual stars may act to hide small-scale features. We note that the region of enhanced
intrinsic reddening in the Tatton et al. (submitted) map covers a significant fraction of the main body of the SMC and thus it is hard to reconcile why this is not
somehow also an inherent feature of the published reddening maps inferred from the RC in the optical\footnote{Preliminary results by Choi et al. (in preparation) using
SMASH $g$- and $i$-band data to map the reddening across the SMC using the colour of RC stars appears to show a similar morphology and reddening levels to those
in Tatton et al. (submitted). We defer further discussion on the potential causes of this discrepancy between this and other optical RC reddening maps to
Choi et al. (in preparation).}.

To place our comparison in a more quantitative context, we note that the median $E(B-V)$ value of our reddening map based on background galaxies with low levels of
intrinsic reddening of $E(B-V)=0.05$\,mag is consistent, to within the uncertainties, with the median values from all the literature reddening maps in Fig.~\ref{fig:red_comp_single}.
We find a minimum difference of $\Delta E(B-V)_{\mathrm{med}}=0.004$\,mag in the comparison with the \cite{Gorski20} map, whereas a maximum difference of
$\Delta E(B-V)_{\mathrm{med}}=-0.061$\,mag is found in our comparison with the ($K_{\mathrm{s}}, Y-K_{\mathrm{s}}$) \cite{Rubele18} map. Note that we define
$\Delta E(B-V)_{\mathrm{med}}$ as the median reddening of our map minus the median reddening of the literature map, such that positive values imply increased levels
of reddening in our map and negative values increased levels of reddening in the relevant literature map. If we allow for the differences in the depth of the SMC sampled, we
still find that the median reddening in all but the two realisations of the \cite{Rubele18} reddening maps are consistent with the median value of our map. This is likely due to
the extended regions of significant reddening observed in the \cite{Rubele18} maps that cover a larger area of the SMC body than the regions of enhanced reddening
in the other maps (see Fig.~\ref{fig:red_comp_single}).

Our reddening map suggests that the highest levels of intrinsic reddening in the SMC are associated with the south-western region of the main body
[$E(B-V)_{\mathrm{max}}=0.325$\,mag]. Where observed, we check for consistency between the level of enhancement in our map compared to others in
Fig.~\ref{fig:red_comp_single}. A minimum difference of $\Delta E(B-V)=0.061$\,mag is found for the ($K_{\mathrm{s}}, Y-K_{\mathrm{s}}$)
\cite{Rubele18} map, whereas a maximum of $\Delta E(B-V)=0.205$\,mag is found for the \cite{Joshi19} map. For the maps that show no obvious enhancements of
reddening in the main body, the differences are all larger than $\Delta E(B-V)=0.240$\,mag. Despite the clumpy and inhomogeneous nature of dust in the SMC, that
tends to be concentrated in and around star-forming regions, if
we, for the sake of comparison, assume that the stars sample, on average, half the
depth of the SMC, then by multiplying the reddening values of the stellar reddening maps by a factor of 2, we find that the enhancement observed in our reddening map
based on background galaxies is consistent, to within the uncertainties, with the enhancements present in the
\cite{Joshi19}, Tatton et al. (submitted) and the ($K_{\mathrm{s}}, J-K_{\mathrm{s}}$) \cite{Rubele18} maps [$\Delta E(B-V)=0.084$, $-0.060$ and $-0.026$\,mag,
respectively]. In contrast, the ($K_{\mathrm{s}}, Y-K_{\mathrm{s}}$) \cite{Rubele18}
map implies significantly higher levels of reddening [$\Delta E(B-V)=-0.204$\,mag] which is hard to reconcile with what we find using background galaxies. It is clear, however,
that unlike the ($K_{\mathrm{s}}, J-K_{\mathrm{s}}$) reddening map, there are strong tile-to-tile variations that could be due to issues
pertaining to the absolute $Y$-band calibration of the VMC data (see Paper~I).

Of the reddening maps shown in Fig.~\ref{fig:red_comp_single}, perhaps the most like-for-like, in terms of sampling the full column density of dust along a given line of
sight, is the HERITAGE dust map of \cite{Gordon14}. It is therefore interesting to note large discrepancies between that map and ours based on background
galaxies. In terms of morphology, the \cite{Gordon14} map suggests that the main body of the SMC has an enhanced level of reddening with respect to the outskirts.
This level of enhancement, however, is significantly lower than what we observe in our map [cf. $E(B-V)_{\mathrm{med}}=0.008$ and $E(B-V)_{\mathrm{max}}=0.063$\,mag for the
\citealp{Gordon14} dust map with $E(B-V)_{\mathrm{med}}=0.050$ and $E(B-V)_{\mathrm{max}}=0.325$\,mag for our reddening map]. Far-IR observations are excellent
tracers of interstellar dust and the long wavelength coverage ($100-500$\,$\mu$m) ensures an almost complete census of the dust mass of the SMC. Note that
although a larger fraction of the dust luminosity is emitted at shorter mid- and near-IR wavelengths, these wavelengths only probe a small fraction of the total dust
mass and thus their exclusion from the \cite{Gordon14} analysis does not significantly affect the resultant reddening map shown in Fig.~\ref{fig:red_comp_single}.

The main difference between the reddening maps based on background galaxies and the dust emission is that there are regions of high reddening in the former for which
there is seemingly little emission in the latter. Although not obvious from Fig.~\ref{fig:red_comp_single} there is a difference in the dust emission associated with the
south-western and north-eastern regions of the main body of the SMC, however these do not translate into the significant difference we observe in the galaxy-inferred
reddening values\footnote{Note that the study of polycyclic aromatic hydrocarbons (PAHs) as part of the S$^{3}$MC project \citep{Bolatto07,Sandstrom10} finds
significant levels of interstellar reddening [$E(B-V)\sim$\,0.3\,mag] in both the south-western and north-eastern regions of the main body of the SMC.}.
One possible reason for this discrepancy could be the spatial scales probed by the different tracers. 
The use of background galaxies provides the reddening within
a very narrow column, whereas the spatial resolution of the images used to create the dust emission map are $\simeq$\,10\,pc (having transformed the spatial resolution
of all images to that of the 500\,$\mu$m observations; see \citealp{Gordon14}). Thus it is possible that the regions in the background galaxy reddening map with significant
reddening could be due to dust clouds with very small spatial scales which are effectively missed by the far-IR observations.

Another potential cause for the discrepancy comes from the assumed far-IR emissivity used to convert the observed far-IR surface brightness into a dust surface density
as well as the subsequent adopted value for $V$-band extinction efficiency factor, $Q_V$. \cite{Gordon14} calibrate the far-IR emissivity using the far-IR/submillimetre
SED of the diffuse MW ISM, however as noted by \citet[see their appendix~A]{Roman-Duval14}, the random error associated with this calibration translates to a
systematic uncertainty on the dust surface density of 13 per cent. Furthermore, there is an additional systematic uncertainty of 22 per cent on the dust surface density related to the
fact that the dust composition, and hence far-IR emissivity, in the MCs is not necessarily identical to that in the MW. Thus, the total systematic uncertainty on the derived dust
surface density due to the poorly constrained nature of the far-IR emissivity (see also discussions in \citealp{Fanciullo05} and \citealp{Clark19}) equates to 35 per cent.
The extinction efficiency factor is a strong function of grain structure, composition and polarisation with studies indicating that variations in these can result in increases
in the value of $Q_V$ by up to a factor of 3 (\citealp{Gupta05}; \citealp*{Voshchinnikov05}). Given the uncertainties inherent in transforming far-IR dust emission observations
into optical interstellar reddening values, and that to bring the reddening map based on dust emission into agreement with that inferred from background galaxies only requires
a scaling of $\sim$\,5, it is not inconceivable that the two independent tracers could in fact provide consistent results.

The differences among reddening maps are likely highlighting interesting physical properties of the relative distributions of dust and stellar populations. The typical
presumption of a smooth screen of dust is manifestly refuted by the different maps produced by young stars, old stars, and now galaxies. At first glance, it is puzzling that
the two probes that are least likely to be spatially correlated with dust -- old stars and galaxies -- produce reddening maps that are in apparent disagreement. However, here
we may be witnessing the effects of the third dimension. Stars are distributed along the line of sight relative to the dust and so some will be foreground to the dust, and
hence unreddened, while others may lie behind the dust, and hence as reddened as background galaxies along the same line of sight. If the reddening and associated
extinction along a line of sight is sufficiently strong to result in a bias in the stellar sample, then reddening values estimated using stars could be severely underestimated and
result in apparent conflict in the maps from stars and from galaxies. Reddening estimates based on near-IR photometry may suffer less from such a bias, perhaps suggesting
why the Tatton et al. (submitted) map is in better agreement with the galaxy-derived map. A complete treatment of this issue is beyond the scope of the current paper, but we do suggest
an outline of a treatment where one begins with the galaxy-derived extinction map as the measurement of the full column-density of dust along the line of sight. Assuming that
the dust is highly confined to the thin midplane, one could distribute stars in the foreground, within the midplane, and background. Within the midplane one would want to
adjust the degree of correlation between stars and dust, particularly for the younger stars. By then mimicking the various sample selection effects and recreating the derived
reddening maps, one could test whether the variety seen among reddening maps is reproduced and at the same time measure the relative distribution of stars and dust.


\section{Summary}
\label{summary}

In this study we have extended the technique introduced in Paper~I to quantify and map the total intrinsic reddening across $\simeq$\,34\,deg$^{2}$ of the SMC
based on the analysis of SEDs of background galaxies. Below we provide a brief overview of the steps involved and our main conclusions.

\begin{enumerate}

\item We use a combination of colour--magnitude and morphological selections to select an initial sample of $\sim$\,500,000 likely background sources from the VMC
PSF photometric catalogues covering the combined SMASH-VMC footprint of the SMC. We consistently measure fluxes for each source using \textsc{lambdar}
and create SEDs from the optical ($ugriz$; SMASH) and near-IR ($YJK_{\mathrm{s}}$; VMC) broadband images

\item We use the \textsc{lephare} $\chi^{2}$ template-fitting code to fit a combination of galaxy, QSO and stellar templates to the SEDs. From these, we select a 
cleaned subsample of 29,274 galaxies with low levels of intrinsic reddening to create an intrinsic reddening map of the SMC. Our reddening
map shows statistically significant levels of enhanced reddening associated with the main body of the SMC compared with negligible levels of reddening in the outskirts
[$\Delta E(B-V) \simeq 0.3$\,mag].

\item We find very good agreement between the calculated \textsc{lephare} photometric redshifts and the spectroscopically determined redshifts for a sample of
46 QSOs behind the SMC (median $\Delta z=-0.006$).

\item We perform a comparison of our intrinsic reddening map with publicly available reddening maps of the SMC. As noted in Paper~I, there is significant
variation amongst the various literature maps that tends to be dependent upon the adopted tracer. Having accounted for the difference in the SMC depth
probed using stellar tracers and background galaxies, we find that the levels of enhanced reddening observed in reddening maps that include
contributions from young stars is consistent with the levels of enhancement we observe in our reddening map. For reddening maps based on older stellar populations
we find a significant discrepancy between the levels of reddening associated with the main body of the SMC. One notable exception to this is the near-IR RC map of
Tatton et al. (submitted) that is consistent with our findings. Our comparison with the HERITAGE dust map of
\cite{Gordon14} also shows a significant discrepancy, however this could be due to the different length scales probed by the respective tracers as well as uncertainties in
the far-IR emissivity and the optical properties of the dust grains.

\item In a future work we aim to use publicly available reddening maps of the SMC to test the inferred dust densities by comparing these to the expected
limits on the gas-to-dust ratio to potentially highlight unrealistic dust mass densities (i.e. high reddening values; see e.g. \citealp{Gordon14,Roman-Duval14}).

\end{enumerate}

\section*{Acknowledgements}

This project has received funding from the European Research Council (ERC) under the European Union's Horizon
2020 research and innovation programme (grant agreement no. 682115).
S.R. acknowledges support from the ERC consolidator grant project STARKEY (grant agreement no. 615604).
Y.C. acknowledges support from NSF grant AST 1655677.
DMD acknowledges financial support from the Spanish Ministry for Science, Innovation and Universities and FEDER funds through grant
AYA2016-81065-C2-2, the State Agency for Research of the Spanish MCIU through the ``Centre of Excellence Severo Ochoa'' award for
the Instituto de Astrof{\'i}sica de Andaluc{\'i}a (SEV-2017-0709) and from grant PGC2018-095049-B-C21.
R.R.M. acknowledges partial support from project BASAL AFB-$170002$ as well as FONDECYT project no. 1170364.
SS acknowledges support from the Science and Engineering Research Board, India through a Ramanujan Fellowship.
The authors would like to thank the Cambridge Astronomy Survey Unit (CASU) and the Wide Field Astronomy
Unit (WFAU) in Edinburgh for providing the necessary data products under the support of the Science and Technology
Facilities Council (STFC) in the U.K.
The authors would also like to thank K. Gordon and J. Roman-Duval for discussions related to the use of the
HERITAGE dust maps.
The authors would like to extend their gratitude to the referee, Geoff Clayton, who provided several comments that improved the manuscript.
This study was based on observations made with VISTA at the La Silla Paranal Observatory under programme ID 179.B-2003.
This project used data obtained with the
Dark Energy Camera (DECam), which was constructed by the
Dark Energy Survey (DES) collaboration.
Funding for the DES Projects has been provided by 
the U.S. Department of Energy, 
the U.S. National Science Foundation, 
the Ministry of Science and Education of Spain, 
the STFC, 
the Higher Education Funding Council for England, 
the National Center for Supercomputing Applications at the University of Illinois at Urbana-Champaign, 
the Kavli Institute of Cosmological Physics at the University of Chicago, 
the Center for Cosmology and Astro-Particle Physics at the Ohio State University, 
the Mitchell Institute for Fundamental Physics and Astronomy at Texas A\&M University, 
Financiadora de Estudos e Projetos, Funda{\c c}{\~a}o Carlos Chagas Filho de Amparo {\`a} Pesquisa do Estado do Rio de Janeiro, 
Conselho Nacional de Desenvolvimento Cient{\'i}fico e Tecnol{\'o}gico and the Minist{\'e}rio da Ci{\^e}ncia, Tecnologia e Inovac{\~a}o, 
the Deutsche Forschungsgemeinschaft, 
and the Collaborating Institutions in the Dark Energy Survey. 
The Collaborating Institutions are 
Argonne National Laboratory, 
the University of California at Santa Cruz, 
the University of Cambridge, 
Centro de Investigaciones En{\'e}rgeticas, Medioambientales y Tecnol{\'o}gicas-Madrid, 
the University of Chicago, 
University College London, 
the DES-Brazil Consortium, 
the University of Edinburgh, 
the Eidgen{\"o}ssische Technische Hoch\-schule (ETH) Z{\"u}rich, 
Fermi National Accelerator Laboratory, 
the University of Illinois at Urbana-Champaign, 
the Institut de Ci{\`e}ncies de l'Espai (IEEC/CSIC), 
the Institut de F{\'i}sica d'Altes Energies, 
Lawrence Berkeley National Laboratory, 
the Ludwig-Maximilians Universit{\"a}t M{\"u}nchen and the associated Excellence Cluster Universe, 
the University of Michigan, 
{the} National Optical Astronomy Observatory, 
the University of Nottingham, 
the Ohio State University, 
the University of Pennsylvania, 
the University of Portsmouth, 
SLAC National Accelerator Laboratory, 
Stanford University, 
the University of Sussex, 
and Texas A\&M University.
 Based on
observations at Cerro Tololo Inter-American Observatory,
National Optical Astronomy Observatory (NOAO Prop. ID:
2013A-0411 and 2013B-0440; PI: Nidever), which is operated
by the Association of Universities for Research in Astronomy
(AURA) under a cooperative agreement with the National
Science Foundation.
Finally, this project has made extensive use of the Tool for OPerations on Catalogues And
Tables (TOPCAT) software package \citep{Taylor05} as well as the following open-source
\texttt{Python} packages: Astropy \citep{Astropy18}, matplotlib \citep{Hunter07}, NumPy
\citep{Oliphant15}, pandas \citep{McKinney10}, and SciPy \citep{Jones01}.

\section*{Data Availability }

The data underlying this article are available in the article and in its online supplementary material.
The SMASH and VMC survey images used to create the galaxy SEDs are publicly available at
\url{https://datalab.noao.edu/smash/smash.php#SecondtDataRelease} and \url{https://www.eso.org/sci/publications/announcements/sciann17313.html},
respectively. The galaxy SEDs will be shared on reasonable request to the corresponding author.

\bibliographystyle{mn3e}
\bibliography{references}

\begin{thebibliography}{68}
\expandafter\ifx\csname natexlab\endcsname\relax\def\natexlab#1{#1}\fi

\bibitem[{{Arnouts} {et~al.}(2007){Arnouts}, {Walcher}, {Le F{\`e}vre},
  {et~al.}}]{Arnouts07}
{Arnouts} S., {Walcher} C.~J., {Le F{\`e}vre} O., {et~al.}, 2007, \aap, 476,
  137

\bibitem[{{Bailey} {et~al.}(2015){Bailey}, {van Loon}, {Sarre}, \&
  {Beckman}}]{Bailey15}
{Bailey} M., {van Loon} J.~T., {Sarre} P.~J., {Beckman} J.~E., 2015, \mnras,
  454, 4013

\bibitem[{{Bell} {et~al.}(2019){Bell}, {Cioni}, {Wright}, {et~al.}}]{Bell19}
{Bell} C. P.~M., {Cioni} M.-R.~L., {Wright} A.~H., {et~al.}, 2019, \mnras, 489,
  3200

\bibitem[{{Bolatto} {et~al.}(2007){Bolatto}, {Simon}, {Stanimirovi{\'c}}, {van
  Loon}, {Shah}, {Venn}, {Leroy}, {Sandstrom}, {Jackson}, {Israel}, {Li},
  {Staveley-Smith}, {Bot}, {Boulanger}, \& {Rubio}}]{Bolatto07}
{Bolatto} A.~D., {Simon} J.~D., {Stanimirovi{\'c}} S., {van Loon} J.~T., {Shah}
  R.~Y., {Venn} K., {Leroy} A.~K., {Sandstrom} K., {Jackson} J.~M., {Israel}
  F.~P., {Li} A., {Staveley-Smith} L., {Bot} C., {Boulanger} F., {Rubio} M.,
  2007, \apj, 655, 212

\bibitem[{{Bouchet} {et~al.}(1985){Bouchet}, {Lequeux}, {Maurice}, {Prevot}, \&
  {Prevot-Burnichon}}]{Bouchet85}
{Bouchet} P., {Lequeux} J., {Maurice} E., {Prevot} L., {Prevot-Burnichon}
  M.~L., 1985, \aap, 149, 330

\bibitem[{{Cardelli} {et~al.}(1989){Cardelli}, {Clayton}, \&
  {Mathis}}]{Cardelli89}
{Cardelli} J.~A., {Clayton} G.~C., {Mathis} J.~S., 1989, \apj, 345, 245

\bibitem[{{Chlewicki}(1985)}]{Chlewicki85}
{Chlewicki} G., 1985, PhD thesis, Observational Constraints on Multimodal
  Interstellar Grain Populations (Leiden University)

\bibitem[{{Choi} {et~al.}(2018){Choi}, {Nidever}, {Olsen}, {et~al.}}]{Choi18}
{Choi} Y., {Nidever} D.~L., {Olsen} K., {et~al.}, 2018, \apj, 866, 90

\bibitem[{{Cioni} {et~al.}(2011){Cioni}, {Clementini}, {Girardi},
  {et~al.}}]{Cioni11}
{Cioni} M.-R.~L., {Clementini} G., {Girardi} L., {et~al.}, 2011, \aap, 527,
  A116

\bibitem[{{Clark} {et~al.}(2019){Clark}, {De Vis}, {Baes}, {Bianchi},
  {Casasola}, {Cassar{\`a}}, {Davies}, {Dobbels}, {Lianou}, {De Looze},
  {Evans}, {Galametz}, {Galliano}, {Jones}, {Madden}, {Mosenkov}, {Verstocken},
  {Viaene}, {Xilouris}, \& {Ysard}}]{Clark19}
{Clark} C.~J.~R., {De Vis} P., {Baes} M., {Bianchi} S., {Casasola} V.,
  {Cassar{\`a}} L.~P., {Davies} J.~I., {Dobbels} W., {Lianou} S., {De Looze}
  I., {Evans} R., {Galametz} M., {Galliano} F., {Jones} A.~P., {Madden} S.~C.,
  {Mosenkov} A.~V., {Verstocken} S., {Viaene} S., {Xilouris} E.~M., {Ysard} N.,
  2019, \mnras, 489, 5256

\bibitem[{{Dahlen} {et~al.}(2013){Dahlen}, {Mobasher}, {Faber},
  {et~al.}}]{Dahlen13}
{Dahlen} T., {Mobasher} B., {Faber} S.~M., {et~al.}, 2013, \apj, 775, 93

\bibitem[{{de Grijs} \& {Bono}(2015)}]{deGrijs15}
{de Grijs} R., {Bono} G., 2015, \aj, 149, 179

\bibitem[{{de Grijs} {et~al.}(2014){de Grijs}, {Wicker}, \& {Bono}}]{deGrijs14}
{de Grijs} R., {Wicker} J.~E., {Bono} G., 2014, \aj, 147, 122

\bibitem[{{Dobrzycki} {et~al.}(2003{\natexlab{a}}){Dobrzycki}, {Macri},
  {Stanek}, \& {Groot}}]{Dobrzycki03a}
{Dobrzycki} A., {Macri} L.~M., {Stanek} K.~Z., {Groot} P.~J.,
  2003{\natexlab{a}}, \aj, 125, 1330

\bibitem[{{Dobrzycki} {et~al.}(2003{\natexlab{b}}){Dobrzycki}, {Stanek},
  {Macri}, \& {Groot}}]{Dobrzycki03b}
{Dobrzycki} A., {Stanek} K.~Z., {Macri} L.~M., {Groot} P.~J.,
  2003{\natexlab{b}}, \aj, 126, 734

\bibitem[{{El Youssoufi} {et~al.}(2019){El Youssoufi}, {Cioni}, {Bell},
  {Rubele}, {Bekki}, {de Grijs}, {Girardi}, {Ivanov}, {Matijevic},
  {Niederhofer}, {Oliveira}, {Ripepi}, {Subramanian}, \& {van
  Loon}}]{ElYoussoufi19}
{El Youssoufi} D., {Cioni} M.-R.~L., {Bell} C. P.~M., {Rubele} S., {Bekki} K.,
  {de Grijs} R., {Girardi} L., {Ivanov} V.~D., {Matijevic} G., {Niederhofer}
  F., {Oliveira} J.~M., {Ripepi} V., {Subramanian} S., {van Loon} J.~T., 2019,
  \mnras, 490, 1076

\bibitem[{{Fanciullo} {et~al.}(2015){Fanciullo}, {Guillet}, {Aniano}, {Jones},
  {Ysard}, {Miville-Desch{\^e}nes}, {Boulanger}, \& {K{\"o}hler}}]{Fanciullo05}
{Fanciullo} L., {Guillet} V., {Aniano} G., {Jones} A.~P., {Ysard} N.,
  {Miville-Desch{\^e}nes} M.~A., {Boulanger} F., {K{\"o}hler} M., 2015, \aap,
  580, A136

\bibitem[{{Freedman} {et~al.}(2020){Freedman}, {Madore}, {Hoyt}, {Jang},
  {Beaton}, {Lee}, {Monson}, {Neeley}, \& {Rich}}]{Freedman20}
{Freedman} W.~L., {Madore} B.~F., {Hoyt} T., {Jang} I.~S., {Beaton} R., {Lee}
  M.~G., {Monson} A., {Neeley} J., {Rich} J., 2020, \apj, 891, 57

\bibitem[{{Geha} {et~al.}(2003){Geha}, {Alcock}, {Allsman}, {et~al.}}]{Geha03}
{Geha} M., {Alcock} C., {Allsman} R.~A., {et~al.}, 2003, \aj, 125, 1

\bibitem[{{Gieren} {et~al.}(2018){Gieren}, {Storm}, {Konorski}, {G{\'o}rski},
  {Pilecki}, {Thompson}, {Pietrzy{\'n}ski}, {Graczyk}, {Barnes}, {Fouqu{\'e}},
  {Nardetto}, {Gallenne}, {Karczmarek}, {Suchomska}, {Wielg{\'o}rski},
  {Taormina}, \& {Zgirski}}]{Gieren18}
{Gieren} W., {Storm} J., {Konorski} P., {G{\'o}rski} M., {Pilecki} B.,
  {Thompson} I., {Pietrzy{\'n}ski} G., {Graczyk} D., {Barnes} T.~G.,
  {Fouqu{\'e}} P., {Nardetto} N., {Gallenne} A., {Karczmarek} P., {Suchomska}
  K., {Wielg{\'o}rski} P., {Taormina} M., {Zgirski} B., 2018, \aap, 620, A99

\bibitem[{{Gordon} {et~al.}(2014){Gordon}, {Roman-Duval}, {Bot},
  {et~al.}}]{Gordon14}
{Gordon} K.~D., {Roman-Duval} J., {Bot} C., {et~al.}, 2014, \apj, 797, 85

\bibitem[{{G{\'o}rski} {et~al.}(2020){G{\'o}rski}, {Zgirski},
  {Pietrzy{\'n}ski}, {Gieren}, {Wielg{\'o}rski}, {Graczyk}, {Kudritzki},
  {Pilecki}, {Narloch}, {Karczmarek}, {Suchomska}, \& {Taormina}}]{Gorski20}
{G{\'o}rski} M., {Zgirski} B., {Pietrzy{\'n}ski} G., {Gieren} W.,
  {Wielg{\'o}rski} P., {Graczyk} D., {Kudritzki} R.-P., {Pilecki} B., {Narloch}
  W., {Karczmarek} P., {Suchomska} K., {Taormina} M., 2020, \apj, 889, 179

\bibitem[{{Gouliermis} {et~al.}(2014){Gouliermis}, {Hony}, \&
  {Klessen}}]{Gouliermis14}
{Gouliermis} D.~A., {Hony} S., {Klessen} R.~S., 2014, \mnras, 439, 3775

\bibitem[{{Groenewegen}(2018)}]{Groenewegen18}
{Groenewegen} M.~A.~T., 2018, \aap, 619, A8

\bibitem[{{Gupta} {et~al.}(2005){Gupta}, {Mukai}, {Vaidya}, {Sen}, \&
  {Okada}}]{Gupta05}
{Gupta} R., {Mukai} T., {Vaidya} D.~B., {Sen} A.~K., {Okada} Y., 2005, \aap,
  441, 555

\bibitem[{{Haschke} {et~al.}(2011){Haschke}, {Grebel}, \& {Duffau}}]{Haschke11}
{Haschke} R., {Grebel} E.~K., {Duffau} S., 2011, \aj, 141, 158

\bibitem[{{Hunter}(2007)}]{Hunter07}
{Hunter} J.~D., 2007, Computing in Science \& Engineering, 9, 90

\bibitem[{{Ilbert} {et~al.}(2009){Ilbert}, {Capak}, {Salvato},
  {et~al.}}]{Ilbert09}
{Ilbert} O., {Capak} P., {Salvato} M., {et~al.}, 2009, \apj, 690, 1236

\bibitem[{{Ilbert} {et~al.}(2013){Ilbert}, {McCracken}, {Le F{\`e}vre},
  {et~al.}}]{Ilbert13}
{Ilbert} O., {McCracken} H.~J., {Le F{\`e}vre} O., {et~al.}, 2013, \aap, 556,
  A55

\bibitem[{{Ivanov} {et~al.}(2016){Ivanov}, {Cioni}, {Bekki}, {de Grijs},
  {Emerson}, {Gibson}, {Kamath}, {van Loon}, {Piatti}, \& {For}}]{Ivanov16}
{Ivanov} V.~D., {Cioni} M.-R.~L., {Bekki} K., {de Grijs} R., {Emerson} J.,
  {Gibson} B.~K., {Kamath} D., {van Loon} J.~T., {Piatti} A.~E., {For} B.-Q.,
  2016, \aap, 588, A93

\bibitem[{{Jarvis} {et~al.}(2013){Jarvis}, {Bonfield}, {Bruce},
  {et~al.}}]{Jarvis13}
{Jarvis} M.~J., {Bonfield} D.~G., {Bruce} V.~A., {et~al.}, 2013, \mnras, 428,
  1281

\bibitem[{{Jones} {et~al.}(2001--){Jones}, {Oliphant}, {Peterson},
  {et~al.}}]{Jones01}
{Jones} E., {Oliphant} T.~E., {Peterson} P., {et~al.}, 2001--, {SciPy}: Open
  source scientific tools for {Python}

\bibitem[{{Joshi} \& {Panchal}(2019)}]{Joshi19}
{Joshi} Y.~C., {Panchal} A., 2019, \aap, 628, A51

\bibitem[{{Kohout} {et~al.}(2014){Kohout}, {Kallonen}, {Suuronen}, {Rochette},
  {Hutzler}, {Gattacceca}, {Badjukov}, {Sk{\'a}la}, {B{\"o}hmov{\'a}}, \&
  {{\v{C}}uda}}]{Kohout14}
{Kohout} T., {Kallonen} A., {Suuronen} J.~P., {Rochette} P., {Hutzler} A.,
  {Gattacceca} J., {Badjukov} D.~D., {Sk{\'a}la} R., {B{\"o}hmov{\'a}} V.,
  {{\v{C}}uda} J., 2014, Meteoritics and Planetary Science, 49, 1157

\bibitem[{{Koz{\l}owski} {et~al.}(2011){Koz{\l}owski}, {Kochanek}, \&
  {Udalski}}]{Kozlowski11}
{Koz{\l}owski} S., {Kochanek} C.~S., {Udalski} A., 2011, \apjs, 194, 22

\bibitem[{{Koz{\l}owski} {et~al.}(2013){Koz{\l}owski}, {Onken}, {Kochanek},
  {Udalski}, {Szyma{\'n}ski}, {Kubiak}, {Pietrzy{\'n}ski}, {Soszy{\'n}ski},
  {Wyrzykowski}, {Ulaczyk}, {Poleski}, {Pietrukowicz}, {Skowron}, {OGLE
  Collaboration}, {Meixner}, \& {Bonanos}}]{Kozlowski13}
{Koz{\l}owski} S., {Onken} C.~A., {Kochanek} C.~S., {Udalski} A.,
  {Szyma{\'n}ski} M.~K., {Kubiak} M., {Pietrzy{\'n}ski} G., {Soszy{\'n}ski} I.,
  {Wyrzykowski} {\L}., {Ulaczyk} K., {Poleski} R., {Pietrukowicz} P., {Skowron}
  J., {OGLE Collaboration}, {Meixner} M., {Bonanos} A.~Z., 2013, \apj, 775, 92

\bibitem[{{Le F{\`e}vre} {et~al.}(2005){Le F{\`e}vre}, {Vettolani}, {Garilli},
  {et~al.}}]{LeFevre05}
{Le F{\`e}vre} O., {Vettolani} G., {Garilli} B., {et~al.}, 2005, \aap, 439, 845

\bibitem[{{Maitra} {et~al.}(2019){Maitra}, {Haberl}, {Ivanov}, {Cioni}, \& {van
  Loon}}]{Maitra19}
{Maitra} C., {Haberl} F., {Ivanov} V.~D., {Cioni} M.-R.~L., {van Loon} J.~T.,
  2019, \aap, 622, A29

\bibitem[{{Mann} \& {Kimura}(2000)}]{Mann00}
{Mann} I., {Kimura} H., 2000, \jgr, 105, 10317

\bibitem[{{McKinney}(2010)}]{McKinney10}
{McKinney} W., 2010, in Proceedings of the 9th Python in Science Conference,
  {van der Walt} S., {Millman} J., eds., p.~51

\bibitem[{{Meixner} {et~al.}(2013){Meixner}, {Panuzzo}, {Roman-Duval},
  {et~al.}}]{Meixner13}
{Meixner} M., {Panuzzo} P., {Roman-Duval} J., {et~al.}, 2013, \aj, 146, 62

\bibitem[{{Muller} {et~al.}(2003){Muller}, {Staveley-Smith}, {Zealey}, \&
  {Stanimirovi{\'c}}}]{Muller03}
{Muller} E., {Staveley-Smith} L., {Zealey} W., {Stanimirovi{\'c}} S., 2003,
  \mnras, 339, 105

\bibitem[{{Muraveva} {et~al.}(2018){Muraveva}, {Subramanian}, {Clementini},
  {Cioni}, {Palmer}, {van Loon}, {Moretti}, {de Grijs}, {Molinaro}, {Ripepi},
  {Marconi}, {Emerson}, \& {Ivanov}}]{Muraveva18}
{Muraveva} T., {Subramanian} S., {Clementini} G., {Cioni} M.-R.~L., {Palmer}
  M., {van Loon} J.~T., {Moretti} M.~I., {de Grijs} R., {Molinaro} R., {Ripepi}
  V., {Marconi} M., {Emerson} J., {Ivanov} V.~D., 2018, \mnras, 473, 3131

\bibitem[{{Nayak} {et~al.}(2018){Nayak}, {Subramaniam}, {Choudhury}, \&
  {Sagar}}]{Nayak18}
{Nayak} P.~K., {Subramaniam} A., {Choudhury} S., {Sagar} R., 2018, \aap, 616,
  A187

\bibitem[{{Nidever} {et~al.}(2017){Nidever}, {Olsen}, {Walker},
  {et~al.}}]{Nidever17}
{Nidever} D.~L., {Olsen} K., {Walker} A.~R., {et~al.}, 2017, \aj, 154, 199

\bibitem[{{Oliphant}(2015)}]{Oliphant15}
{Oliphant} T.~E., 2015, {Guide to NumPy}. CreateSpace Independent Publishing
  Platform, 2nd ed.

\bibitem[{{Paturel} {et~al.}(2003){Paturel}, {Petit}, {Prugniel}, {Theureau},
  {Rousseau}, {Brouty}, {Dubois}, \& {Cambr{\'e}sy}}]{Paturel03}
{Paturel} G., {Petit} C., {Prugniel} P., {Theureau} G., {Rousseau} J., {Brouty}
  M., {Dubois} P., {Cambr{\'e}sy} L., 2003, \aap, 412, 45

\bibitem[{{Prevot} {et~al.}(1984){Prevot}, {Lequeux}, {Prevot}, {Maurice}, \&
  {Rocca-Volmerange}}]{Prevot84}
{Prevot} M.~L., {Lequeux} J., {Prevot} L., {Maurice} E., {Rocca-Volmerange} B.,
  1984, \aap, 132, 389

\bibitem[{{Riess} {et~al.}(2019){Riess}, {Casertano}, {Yuan}, {Macri}, \&
  {Scolnic}}]{Riess19}
{Riess} A.~G., {Casertano} S., {Yuan} W., {Macri} L.~M., {Scolnic} D., 2019,
  \apj, 876, 85

\bibitem[{{Riess} {et~al.}(2009){Riess}, {Macri}, {Casertano}, {Sosey},
  {Lampeitl}, {Ferguson}, {Filippenko}, {Jha}, {Li}, {Chornock}, \&
  {Sarkar}}]{Riess09}
{Riess} A.~G., {Macri} L., {Casertano} S., {Sosey} M., {Lampeitl} H.,
  {Ferguson} H.~C., {Filippenko} A.~V., {Jha} S.~W., {Li} W., {Chornock} R.,
  {Sarkar} D., 2009, \apj, 699, 539

\bibitem[{{Ripepi} {et~al.}(2017){Ripepi}, {Cioni}, {Moretti},
  {et~al.}}]{Ripepi17}
{Ripepi} V., {Cioni} M.-R.~L., {Moretti} M.~I., {et~al.}, 2017, \mnras, 472,
  808

\bibitem[{{Ripepi} {et~al.}(2019){Ripepi}, {Molinaro}, {Musella}, {Marconi},
  {Leccia}, \& {Eyer}}]{Ripepi19}
{Ripepi} V., {Molinaro} R., {Musella} I., {Marconi} M., {Leccia} S., {Eyer} L.,
  2019, \aap, 625, A14

\bibitem[{{Roman-Duval} {et~al.}(2014){Roman-Duval}, {Gordon}, {Meixner},
  {et~al.}}]{Roman-Duval14}
{Roman-Duval} J., {Gordon} K.~D., {Meixner} M., {et~al.}, 2014, \apj, 797, 86

\bibitem[{{Rubele} {et~al.}(2015){Rubele}, {Girardi}, {Kerber}, {Cioni},
  {Piatti}, {Zaggia}, {Bekki}, {Bressan}, {Clementini}, {de Grijs}, {Emerson},
  {Groenewegen}, {Ivanov}, {Marconi}, {Marigo}, {Moretti}, {Ripepi},
  {Subramanian}, {Tatton}, \& {van Loon}}]{Rubele15}
{Rubele} S., {Girardi} L., {Kerber} L., {Cioni} M.-R.~L., {Piatti} A.~E.,
  {Zaggia} S., {Bekki} K., {Bressan} A., {Clementini} G., {de Grijs} R.,
  {Emerson} J.~P., {Groenewegen} M.~A.~T., {Ivanov} V.~D., {Marconi} M.,
  {Marigo} P., {Moretti} M.-I., {Ripepi} V., {Subramanian} S., {Tatton} B.~L.,
  {van Loon} J.~T., 2015, \mnras, 449, 639

\bibitem[{{Rubele} {et~al.}(2018){Rubele}, {Pastorelli}, {Girardi}, {Cioni},
  {Zaggia}, {Marigo}, {Bekki}, {Bressan}, {Clementini}, {de Grijs}, {Emerson},
  {Groenewegen}, {Ivanov}, {Muraveva}, {Nanni}, {Oliveira}, {Ripepi}, {Sun}, \&
  {van Loon}}]{Rubele18}
{Rubele} S., {Pastorelli} G., {Girardi} L., {Cioni} M.-R.~L., {Zaggia} S.,
  {Marigo} P., {Bekki} K., {Bressan} A., {Clementini} G., {de Grijs} R.,
  {Emerson} J., {Groenewegen} M.~A.~T., {Ivanov} V.~D., {Muraveva} T., {Nanni}
  A., {Oliveira} J.~M., {Ripepi} V., {Sun} N.-C., {van Loon} J.~T., 2018,
  \mnras, 478, 5017

\bibitem[{{Sandstrom} {et~al.}(2010){Sandstrom}, {Bolatto}, {Draine}, {Bot}, \&
  {Stanimirovi{\'c}}}]{Sandstrom10}
{Sandstrom} K.~M., {Bolatto} A.~D., {Draine} B.~T., {Bot} C.,
  {Stanimirovi{\'c}} S., 2010, \apj, 715, 701

\bibitem[{{Smartt} {et~al.}(2015){Smartt}, {Valenti}, {Fraser},
  {et~al.}}]{Smartt15}
{Smartt} S.~J., {Valenti} S., {Fraser} M., {et~al.}, 2015, \aap, 579, A40

\bibitem[{{Stanimirovi{\'c}} {et~al.}(1999){Stanimirovi{\'c}},
  {Staveley-Smith}, {Dickey}, {Sault}, \& {Snowden}}]{Stanimirovic99}
{Stanimirovi{\'c}} S., {Staveley-Smith} L., {Dickey} J.~M., {Sault} R.~J.,
  {Snowden} S.~L., 1999, \mnras, 302, 417

\bibitem[{{Taylor}(2005)}]{Taylor05}
{Taylor} M.~B., 2005, in Astronomical Society of the Pacific Conference Series,
  Vol. 347, Astronomical Data Analysis Software and Systems XIV, {Shopbell} P.,
  {Britton} M., {Ebert} R., eds., p.~29

\bibitem[{{The Astropy Collaboration} {et~al.}(2018){The Astropy
  Collaboration}, {Price-Whelan}, {Sip{\H o}cz}, {G{\"u}nther},
  {et~al.}}]{Astropy18}
{The Astropy Collaboration}, {Price-Whelan} A.~M., {Sip{\H o}cz} B.~M.,
  {G{\"u}nther} H.~M., {et~al.}, 2018, \aj, 156, 123

\bibitem[{{Udalski}(2003)}]{Udalski03}
{Udalski} A., 2003, \actaa, 53, 291

\bibitem[{{Udalski} {et~al.}(2015){Udalski}, {Szyma{\'n}ski}, \&
  {Szyma{\'n}ski}}]{Udalski15}
{Udalski} A., {Szyma{\'n}ski} M.~K., {Szyma{\'n}ski} G., 2015, \actaa, 65, 1

\bibitem[{{V{\'e}ron-Cetty} \& {V{\'e}ron}(2010)}]{Veron-Cetty10}
{V{\'e}ron-Cetty} M.-P., {V{\'e}ron} P., 2010, \aap, 518, A10

\bibitem[{{Voshchinnikov} {et~al.}(2005){Voshchinnikov}, {Il'in}, \&
  {Henning}}]{Voshchinnikov05}
{Voshchinnikov} N.~V., {Il'in} V.~B., {Henning} T., 2005, \aap, 429, 371

\bibitem[{{Whittet}(2003)}]{Whittet03}
{Whittet} D.~C.~B., 2003, {Dust in the Galactic Environment}. Bristol:
  Institute of Physics Publishing, 2nd ed.

\bibitem[{{Wright} {et~al.}(2016){Wright}, {Robotham}, {Bourne},
  {et~al.}}]{Wright16}
{Wright} A.~H., {Robotham} A.~S.~G., {Bourne} N., {et~al.}, 2016, \mnras, 460,
  765

\bibitem[{{Zaritsky} {et~al.}(2002){Zaritsky}, {Harris}, {Thompson}, {Grebel},
  \& {Massey}}]{Zaritsky02}
{Zaritsky} D., {Harris} J., {Thompson} I.~B., {Grebel} E.~K., {Massey} P.,
  2002, \aj, 123, 855

\bibitem[{{Zivkov} {et~al.}(2018){Zivkov}, {Oliveira}, {Petr-Gotzens}, {Cioni},
  {Rubele}, {van Loon}, {Bekki}, {Cusano}, {de Grijs}, {Ivanov}, {Marconi},
  {Niederhofer}, {Ripepi}, \& {Sun}}]{Zivkov18}
{Zivkov} V., {Oliveira} J.~M., {Petr-Gotzens} M.~G., {Cioni} M.-R.~L., {Rubele}
  S., {van Loon} J.~T., {Bekki} K., {Cusano} F., {de Grijs} R., {Ivanov} V.~D.,
  {Marconi} M., {Niederhofer} F., {Ripepi} V., {Sun} N.-C., 2018, \aap, 620,
  A143

\end{thebibliography}

\end{document}